\documentclass{PoS}

\def \ms {{\overline{\mbox{MS}}}}

\newcommand{\bea}{\begin{eqnarray}}
\newcommand{\eea}{\end{eqnarray}}
\newcommand{\be}{\begin{equation}}
\newcommand{\ee}{\end{equation}}

\newcommand{\Li}{\mathop{\mathrm{Li}}\nolimits}

\title{Small $x$ behavior of parton distributions. Analytical and ``frozen'' 
coupling constants. BFKL corrections }

\ShortTitle{Small x behavior of parton distributions
}

\author{\speaker{Anatoly Kotikov}
         \thanks{The work was supported in part by RFBR grant 
No. 10-02-01259-a.
}\\
        BLThPh, Joint Institute for Nuclear Research, Dubna\\
        E-mail: \email{kotikov@theor.jinr.ru}}

\abstract{It is shown that in 
the
leading twist approximation of the Wilson operator product expansion
with ``frozen'' and analytic 
strong coupling constants, 
Bessel-inspired
behavior of 
the structure functions $F_2$ and $F_2^{cc}$
and also 
the  derivative $\partial \ln F_2/\partial \ln(1/x)$
at small $x$ values,
obtained for
a flat initial condition in the DGLAP evolution equations,
leads to 
good agreement with the 
deep inelastic 
scattering  experimental data from HERA.}

\FullConference{XXI International Baldin Seminar on High Energy Physics Problems \\
                 September 10-15, 2012\\
                 JINR, Dubna, Russia}

\begin{document}

\section{Introduction}

The experimental data from
HERA on the deep-inelastic scattering (DIS) structure function
(SF) $F_2$ \cite{H197}-\cite{Aaron:2009aa},
%\cite{H197,ZEUS01,Aaron:2009aa},
%and 
its derivative
%s $\partial F_2/\partial \ln(Q^2)$ \cite{H197,Surrow} and 
$\partial \ln F_2/\partial \ln(1/x)$ \cite{Surrow}-\cite{DIS02} 
%\cite{H1slo,Surrow,DIS02} 
and the heavy quark parts $F_2^{cc}$ and $F_2^{bb}$ \cite{Collaboration:2009jy}-\cite{Lipka:2009zza}
enable us to enter into
a very interesting kinematical range for
testing the theoretical ideas on the behavior of quarks and gluons carrying
a very low fraction of momentum of the proton, the so-called small-$x$
region. In this limit one expects that 
the conventional treatment based on the
Dokshitzer--Gribov--Lipatov--Altarelli--Parisi (DGLAP) equations \cite{DGLAP}
does not account for contributions to the cross section which are
leading in $\alpha_s \ln(1/x)$ and, moreover, the parton distribution
function (PDFs),
%densities (PD), 
in particular the gluon ones, are becoming large and need to develop a 
high density formulation of QCD.

 However, the
reasonable agreement between HERA data and the next-to-leading-order (NLO) approximation of
perturbative
QCD has been observed for $Q^2 \geq 2$ GeV$^2$ (see reviews in \cite{CoDeRo}
and references therein) and, thus,
perturbative QCD could describe the
evolution of $F_2$ and its derivatives
%structure functions
up to very low $Q^2$ values,
traditionally explained by soft processes.
%It is of fundamental importance to find out the kinematical region where
%the well-established perturbative QCD formalism
%can be safely applied at small $x$.

The standard program to study the $x$ behavior of
quarks and gluons
is carried out by comparison of data
with the numerical solution of the DGLAP
%Dokshitzer-Gribov-Lipatov-Altarelli-Parisi (DGLAP)
equation
\cite{DGLAP}\footnote{ At small $x$ there is another approach
based on the Balitsky--Fadin--Kuraev--Lipatov (BFKL) equation 
\cite{BFKL}, whose application will be dicussed below in Appendix A.
%is out of the scope of this work. 
} by
fitting the parameters of the PDF
$x$-profile 
%of partons 
at some initial $Q_0^2$ and
the QCD energy scale $\Lambda$ \cite{fits}-\cite{Kotikov:2010bm}.
%\cite{fits,GRV,Ourfits} [KKS].
However, for analyzing exclusively the
%small
low-$x$ region, there is the alternative of doing a simpler analysis
by using some of the existing analytical solutions of DGLAP evolution
in the 
%small
low-$x$
limit \cite{BF1}--\cite{HT}.
This was done so in \cite{BF1}
where it was pointed out that the HERA small-$x$ data can be
interpreted in 
terms of the so-called doubled asymptotic scaling (DAS) phenomenon
related to the asymptotic 
behavior of the DGLAP evolution 
discovered many years ago \cite{Rujula}.

The study of \cite{BF1} was extended in \cite{Munich,Q2evo,HT}
to include the finite parts of anomalous dimensions
of Wilson operators
% and Wilson coefficients
\footnote{ 
In the standard DAS approximation \cite{Rujula} only the singular
parts of the anomalous dimensions were used.}.
This has led to predictions \cite{Q2evo,HT} of the small-$x$ asymptotic PDF
form 
%of parton distributions (PD)
in the framework of the DGLAP dynamics
%equation 
starting at some $Q^2_0$ with
the flat function
 \begin{eqnarray}
f_a (Q^2_0) ~=~
A_a ~~~~(\mbox{hereafter } a=q,g), \label{1}
 \end{eqnarray}
where $f_a$ are the parton distributions multiplied by $x$
and $A_a$ are unknown parameters to be determined from the data.

%From now on, we 
We refer to the approach of \cite{Munich,Q2evo,HT} as
{\it generalized} DAS approximation. In that approach
%generalized DAS 
the flat initial conditions in Eq. (\ref{1}) determine the
basic role of the singular parts of anomalous dimensions,
as in the standard DAS case, while
the contribution from finite parts 
of anomalous dimensions and from Wilson coefficients can be
considered as corrections which are, however, important for better 
agreement with experimental data.
In the present paper, similary to
\cite{BF1}--\cite{HT}, we neglect
the contribution from the non-singlet quark component.

The use of the flat initial condition given in Eq. (\ref{1}) is
supported by the actual experimental situation: low-$Q^2$ data
\cite{NMC,H197,lowQ2,Surrow} are well described for $Q^2 \leq 0.4$ GeV$^2$
by Regge theory with Pomeron intercept
$\alpha_P(0) \equiv \lambda_P +1 =1.08$,
closed to the standard ($\alpha_P(0) =1$) one. 
The small rise of HERA data \cite{H197,Surrow,lowQ2,lowQ2N}
at low $Q^2$ can be 
%naturally 
explained, for example, by contributions of
%including
higher twist operators
%terms 
(see \cite{HT}).
%Moreover, HERA data \cite{H197,H101,lowQ2,lowQ2N}
%with $Q^2 > 1$ GeV$^2$ are in good agreement with predictions from
%the GRV parton densities \cite{GRV}, 
%which 
%are very closed, at least conceptually, to the generalized DAS approach.\\

The purpose of this paper is to demostrate a good agreement between 
the predictions from the generalized DAS approach and the HERA experimental data
\cite{H197,ZEUS01} and \cite{Collaboration:2009jy}-\cite{Lipka:2009zza}
%[HEAVY] (see Figs. 1--3) 
for SF $F_2$ and $F_2^{cc}$  and also
to compare the predictions 
%from the generalized DAS approach 
for
%the small $Q^2$ behavior of 
the slope $\partial  \ln F_2/\partial \ln (1/x)$ 
%of the structure function $F_2$
with the 
%modern 
H1 and ZEUS 
%experimental 
data \cite{H1slo,Surrow,DIS02} (see Figs. 1--8).
%[H1slope,ZEUSslope].
%Very recently new precise experimental data on $\lambda (Q^2)$ 
%has become available \cite{H1slo}.
Looking at the H1 data points \cite{H197} shown in Figs. 5, 6 and 7 one can conclude 
that $\lambda (Q^2)$
is independent of $x$ within the experimental uncertainties
for fixed $Q^2$ in the range $x <0.01$. Indeed, the data are well
described by the power behavior
\begin{eqnarray}
F_2 (x,Q^2) ~=~ C x^{-\lambda (Q^2)},
\label{1dd}
\end{eqnarray}
where $\lambda (Q^2) = \hat a \ln(Q^2/\Lambda^2)$ with $C \approx 0.18,
\hat a \approx 0.048$ and $ \Lambda =292$ MeV \cite{H1slo}.
The linear rise of the exponent $\lambda (Q^2)$ with $\ln Q^2$ is also 
explicitly shown in Figs. 5, 6 and 7 by the dashed line.
%curve.

The
rise of $\lambda (Q^2)$ linearly with $\ln Q^2$ can be tracted in strong 
nonperturbative way (see \cite{Schrempp} and references therein), i.e.,
$\lambda (Q^2) \sim 1/\alpha_s(Q^2)$. The previous analysis \cite{KoPa02}, 
however, demonstrated
that the rise can be explained naturally in the framework of 
perturbative QCD.

%A similar analysis extracting $\lambda (Q^2)$ as a function of $x$
%has been carried out by the ZEUS Collaboration. 
%This behavior can also be inferred
%\footnote{ 
%These points lie slightly below the corresponding ZEUS data but all the
%results are in agreement within modern experimental errors.}
%from Fig. 2.\\

The ZEUS and H1 Collaborations have also presented \cite{Surrow,DIS02} 
%several
new preliminary data
%points 
for $\lambda (Q^2)$ at quite low values of $Q^2$.
As it
is possible to see in Fig. 8 of  \cite{Surrow}, the ZEUS value for 
$\lambda (Q^2)$ is consistent with a constant $\sim 0.1$ at $Q^2 <
0.6$ GeV$^2$, as it is expected under the assumption of single soft Pomeron
exchange within the framework of Regge phenomenology. 
These points lie slightly below the corresponding ZEUS data but all the
results are in agreement within modern experimental errors.

It is important to extend the analysis of \cite{Q2evo,HT,KoPa02} to 
low $Q^2$ range
with the help of well-known infrared modifications of the strong coupling 
constant. Indeed, in Ref. \cite{Cvetic:2009kw}, we have used
%will use 
the ``frozen'' and analytic versions (see,  
\cite{Greco} and \cite{ShiSo}, respectively).

This contribution
%The paper 
is organized as follows.  Sections 2 and 3 contain 
basic formulae,
%for the slope $d \ln F_2/d\ln (1/x)$ in generalized DAS
%%double-logarithmic approximation, 
which are
needed for the present study and were
%that were
previously obtained in \cite{Q2evo,HT,KoPa02,Cvetic:2009kw,Illarionov:2008be}.
In Sections 4 and 5 we compare our calculations with H1 and ZEUS
%$d\ln F_2/d\ln(1/x)$
experimental data and present
%discuss 
the obtained results.
Some discussions can be found in the conclusions.
Some preliminary results accounting for BFKL corrections in our analysis can be found 
in Appendix A.
It is hoped that the inclusion of these corrections will improve the agreement with the experimantal 
data for $F_2$ and its slope at $Q^2 \sim 1\div 2$ Gev$^2$.

\section{
Generalized DAS
approach} \indent

The flat initial condition (\ref{1}) corresponds to the case when parton density
%distributions
tend  to some constant value at $x \to 0$ and at some initial value $Q^2_0$.
%(\ref{1}).
The main ingredients of the results \cite{Q2evo,HT}, are:
\begin{itemize}
\item
Both, the gluon and quark singlet densities are presented in terms of two
components ($"+"$ and $"-"$) which are obtained from the analytic 
$Q^2$-dependent expressions of the corresponding ($"+"$ and $"-"$) PDF
%parton distributions 
moments.
\footnote{Such an approach has been developed  \cite{Albino:2011si}
recently also for the fragmentation function, 
whose first moments (ie mean multiplicities of quarks and gluons) were analyzed
\cite{Bolzoni:2012ii}. The results are 
in good agreement with the experimental data
(see contribution \cite{Bolzoni:2012cv} by Paolo Bolzoni to this Proceedings).}
\item
The twist-two part of the $"-"$ component is constant at small $x$ at any 
values of $Q^2$,
whereas the one of the $"+"$ component grows at $Q^2 \geq Q^2_0$ as
\begin{equation}
\sim e^{\sigma},~~~
%\exp{\sigma},~~~
\sigma = 2\sqrt{\left[ \left|\hat{d}_+\right| s
%\ln \left( \frac{a_s(Q^2_0)}{a_s(Q^2)} \right) 
- \left( \hat{d}_{++}
%\hat{D}_+ 
+  \left|\hat{d}_+\right|
%\hat{d}_+ 
\frac{\beta_1}{\beta_0} \right) p
% \Bigl( a_s(Q^2_0) - a_s(Q^2) \Bigr)
\right] \ln \left( \frac{1}{x} \right)}  \ ,~~~ \rho=\frac{\sigma}{2\ln(1/x)} \ ,
\label{intro:1}
\end{equation}
where $\sigma$ and $\rho$
%$=\sigma/(2\ln(1/x))$ 
are the generalized Ball--Forte
variables,
\begin{equation}
s=\ln \left( \frac{a_s(Q^2_0)}{a_s(Q^2)} \right),~~
p= a_s(Q^2_0) - a_s(Q^2),~~~
\hat{d}_+ = - \frac{12}{\beta_0},~~~
%\hat{D}_+ 
\hat{d}_{++} =  \frac{412}{27\beta_0}.
\label{intro:1a}
\end{equation}
%$8[23 C_A - 26 C_F]T_Rf/(9\beta_0)$.
%%\item
%%The recently observed difference between small $x$ behavior of sea
%%quark and gluon densities at $Q^2=Q^2_0$ are incorporated (in [IKP])
%%by high-twist corrections to the above twist-two approximation.
\end{itemize}
Hereafter we use the notation
$a_s=\alpha_s/(4\pi)$.
The first two coefficients of the QCD $\beta$-function in the $\ms$-scheme
are $\beta_0 = 11 -(2/3) f$
%$(11/3) C_A - (4/3) T_R f$ 
and $\beta_1 =  102 -(114/9) f$
%$(2/3)[17 C_A^2 - 10 C_A T_R f - 6 C_F T_R f]$ 
with $f$ is being the number of active quark flavors.
%This new presentation as a function of the
%$SU(N)$ group casimirs, with $f$ active flavors, $C_A = N$, $T_R = 1/2$,
%$T_F = T_R f$ and  $C_F = (N^2 - 1)/(2N)$ permits to apply our results
%to,  for example, the popular $N=1$ supersymmetric model.
%Of course, for $N=3$ one has the QCD result.
% \cite{Kotikov:1999}.

Note here that the perturbative coupling constant $a_s(Q^2)$ is different at
the leading-order (LO) and NLO approximations. Indeed, from the renormalization group equation
we can obtain the following equations for the coupling constant
%\begin{subequations}
%\label{as:LO&NLO}
\begin{eqnarray}
 \frac{1}{a_s^{\rm LO}(Q^2)} \, = \, \beta_0 
 \ln{\left(\frac{Q^2}{\Lambda^2_{\rm LO}}\right)}
\label{as:LO} 
\end{eqnarray}
at the LO approximation and
\begin{eqnarray}
 \frac{1}{a_s(Q^2)} \, + \,
 \frac{\beta_1}{\beta_0} \ln{\left[
 \frac{\beta_0^2 a_s(Q^2)}{\beta_0+ \beta_1 a_s(Q^2)}\right]} \, = \, 
 \beta_0 \ln{\left(\frac{Q^2}{\Lambda^2}\right)}
\label{as:NLO}
\end{eqnarray}
at the NLO approximation.
%\end{subequations}
Usually at the NLO level ${\rm \overline{MS}}$-scheme is used, so we apply
$\Lambda = \Lambda_{\rm \overline{MS}}$ below.
%in the Eqs.~(\ref{an:NLO}) and (\ref{as:NLO}).

\subsection{Parton distributions and the structure function $F_2$
%$Q^2$ dependence of the slope $d \ln F_2/d\ln (1/x)$ in g
%Generalized DAS approach
%double-logarithmic approximation
} 

%Here, for simplicity we consider only  the LO
%%leading order (LO) 
%approximation\footnote{
%The NLO results may be found in  \cite{Q2evo,HT}.}.
The results for parton densities and $F_2$
%of Refs. \cite{Q2evo,HT} 
are following:
\begin{itemize}
\item
The structure function $F_2$ has the form:
%Both, the gluon and quark singlet densities are presented in terms of two
%components ($'+'$ and $'-'$) 
\begin{eqnarray}
	F_{2,{\rm LO}}(x,Q^2) &=& e \, f_{q,,{\rm LO}}(x,Q^2),
%\label{r10} 
\nonumber \\
	f_{a,{\rm LO}}(x,Q^2) &=& f_{a,{\rm LO}}^{+}(x,Q^2) + 
f_{a,{\rm LO}}^{-}(x,Q^2)
\label{8a}
\end{eqnarray}
at the LO approximation, where
%which are obtained from the analytic 
%$Q^2$-dependent expressions of the corresponding ($'+'$ and $'-'$) PDF
%%parton distributions 
%moments. Here, 
\be e=(\sum_1^f e_i^2)/f  \label{8aa}
\ee
is the average charge square, and \\
%and $f$ is the number of active quark flavors.\\
\begin{eqnarray}
	F_2(x,Q^2) &=& e \, \left( f_q(x,Q^2) + \frac{2}{3} f a_s(Q^2)
f_g(x,Q^2)\right),
%\label{r10} 
\nonumber \\
	f_a(x,Q^2) &=& f_a^{+}(x,Q^2) + f_a^{-}(x,Q^2)
\label{8ab}
\end{eqnarray}
at the NLO approximation.

\item
The small-$x$ asymptotic results for the LO parton densities $f^{\pm}_{a,{\rm LO}}$ are
\begin{eqnarray}
	f^{+}_{g,{\rm LO}}(x,Q^2) &=& \biggl(A_g + \frac{4}{9} A_q \biggl)
		\tilde I_0(\sigma_{\rm LO}) \; e^{-\overline d_{+} s_{\rm LO}} 
~+~ O(\rho_{\rm LO}),
	\label{8.0} \\
	f^{+}_{q,{\rm LO}}(x,Q^2) &=& 
\frac{f}{9}\biggl(A_g + \frac{4}{9} A_q \biggl)
		\rho_{\rm LO} \tilde I_1(\sigma_{\rm LO}) \; e^{-\overline d_{+} s_{\rm LO}} 
~+~ O(\rho),
	\label{8.01} \\
	f^{-}_{g,{\rm LO}}(x,Q^2) &=& -\frac{4}{9} A_q e^{- d_{-} s_{\rm LO}} ~+~ O(x),
	\label{8.00} \\
	f^{-}_{q,{\rm LO}}(x,Q^2) &=& A_q e^{-d_{-} s_{\rm LO}} ~+~ O(x),
	\label{8.02}
\end{eqnarray}
where
\be \overline d_{+} = 1 + 20f/(27\beta_0),~~
%$ and          
d_{-} = 16f/(27\beta_0) \label{8.02a1} 
\ee
are the regular parts of the anomalous dimensions $d_{+}(n)$ and $d_{-}(n)$, 
respectively, in the limit $n\to1$\footnote{
We denote the singular and regular parts of a given quantity $k(n)$ in the
limit $n\to1$ by $\hat k/(n-1)$ and $\overline k$, respectively.}.
%Here $n$ is the variable of the PD Mellin transform.
Here $n$ is
the variable in Mellin space.
%
%We define the variable
% \begin{eqnarray}
%s=ln\left(\frac{a_s(Q^2_0)}{a_s(Q^2)}\right)
%\label{2.4}
% \end{eqnarray}
%
The functions $\tilde I_{\nu}$ ($\nu=0,1$) 
%in Eqs. (\ref{8.0,8.01}) 
are related to the modified Bessel
function $I_{\nu}$
and to the Bessel function $J_{\nu}$ by:
\begin{equation}
\tilde I_{\nu}(\sigma) =
\left\{
\begin{array}{ll}
I_{\nu}(\sigma), & \mbox{ if } s \geq 0 \\
i^{-\nu} J_{\nu}(i\sigma), \ i^2=-1, \ & \mbox{ if } s \leq 0 
\end{array}
\right. .
\label{4}
\end{equation}
At the LO, 
the variables 
%$s$, 
$\sigma_{\rm LO}$ and $\rho_{\rm LO}$ are
%argument $\sigma$ is 
given by Eq. (\ref{intro:1}) when $p=0$, i.e.
\begin{equation}
\sigma_{\rm LO} = 2\sqrt{\left|\hat{d}_+\right| s_{\rm LO}
 \ln \left( \frac{1}{x} \right)}  \ ,~~~ \rho_{\rm LO}=\frac{\sigma_{\rm LO}}{2\ln(1/x)} \ ,
\label{intro:1b}
\end{equation}
and the variable $s_{\rm LO}$ is given by Eq. (\ref{intro:1a}) with $a_s^{\rm LO}(Q^2)$
 as in Eq. (\ref{as:LO}).

\item
The small-$x$ asymptotic results for the NLO parton densities $f^{\pm}_a$ are
\begin{eqnarray}
	f^{+}_g(x,Q^2) &=& A_g^+(Q^2,Q_0^2)
%\biggl(A_g + \frac{4}{9} A_q \biggl)
		\tilde I_0(\sigma) \; e^{-\overline d_{+} s -\overline D_{+} p} ~+~ O(\rho),
	\label{8.0A} \\
	f^{+}_q(x,Q^2) &=& A_q^+
%\frac{f}{9}\biggl(A_g + \frac{4}{9} A_q \biggl) 
\left[\left(1-\overline d_{+-}^q a_s(Q^2)\right)
		\rho \tilde I_1(\sigma) + 20 a_s(Q^2) I_0(\sigma) \right]
\; e^{-\overline d_{+}(1) s -\overline D_{+} p} ~+~ O(\rho),~~~~~~~~
	\label{8.01A} \\
	f^{-}_g(x,Q^2) &=& A_g^-(Q^2,Q_0^2)
%-\frac{4}{9} A_q 
e^{- d_{-}(1) s-D_{-} p} ~+~ O(x),
	\label{8.00A} \\
	f^{-}_q(x,Q^2) &=& A_q^-
%A_q 
e^{-d_{-}(1) s-D_{-} p} ~+~ O(x),
	\label{8.02NLO}
\end{eqnarray}
where
\be
 D_{\pm}=d_{\pm\pm}-\frac{\beta_1}{\beta_0} d_{\pm}
   \label{8.02aa}
\ee
and similar for $\hat D_{+}$ and $\overline D_{+}$,
%\begin{eqnarray}
\bea
A_g^{+}(Q^2,Q_0^2) &=& \left(1-\frac{80f}{81}a_s(Q)\right)A_g + \frac{4}{9}
\left(1+\Bigl(3+\frac{f}{27}\Bigr)a_s(Q_0) -\frac{80f}{81}a_s(Q)\right)A_q,
\nonumber \\
\qquad A_g^{-}(Q^2,Q_0^2) &=& A_g - A_g^{+}(Q^2_0,Q^2) \, .
\label{8.02a}
\eea
%\end{eqnarray}
%and
The coupling constant $a_s(Q^2)$ 
is introduced in Eq. (\ref{as:NLO}). The variables $\hat d_{+}$, $\hat d_{++}$
$\overline d_{+}$ and $d_{-}$ are diven in Eqs. (\ref{intro:1a}) and (\ref{8.02a1}), respectively.
The variables $\overline d_{++}$, $d_{--}$ and $d_{+-}^q$ have the form
\begin{eqnarray}
%a_s(\mu)&=&\frac{\alpha_s(\mu)}{4\pi},\qquad
%\hat d_{+} = - \frac{12}{\beta_0},\qquad
%\overline d_{+}
%%(1) 
%= 1 + \frac{20f}{27\beta_0},\qquad
%d_{-}
%%(1) 
%= \frac{16f}{27\beta_0},\nonumber \\
%\hat d_{++}&=& \frac{412f}{27\beta_0},\qquad
\overline d_{++} &=&
%(1) = 
\frac{8}{\beta_0} \left(
36\zeta_3+33\zeta_2-\frac{1643}{12} + \frac{2f}{9}
\left[\frac{68}{9}-4\zeta_2-\frac{13f}{243}\right]\right), \nonumber \\
d_{--}
%(1) 
&=& \frac{16}{9\beta_0} \left(
2\zeta_3-3\zeta_2+\frac{13}{4} + f
\left[4\zeta_2-\frac{23}{18}+\frac{13f}{243}\right]\right),~~~d_{+-}^q = 23 -12\zeta_2 - \frac{13f}{81},
~~~~~~~
\label{8.02b}
\end{eqnarray}
with $\zeta_3$ and $\zeta_2$ are Eller functions.
%$\beta_0$ and $\beta_1$ are first two coefficients of QCD $\beta$-function.
\footnote{Note that evaluation of the results (\ref{8.0})-(\ref{8.02b})
need the knowledge of the analytic continuation of the anomalous dimansions
and coefficient functions. The analytic continuation can be found in Refs.
\cite{Kazakov:1987jk}. It was used also for the fits 
\cite{Ourfits,Kotikov:2010bm}.}

\end{itemize}

\subsection{Effective slopes}

Contrary to the approach in  \cite{BF1}-\cite{HT}
%\cite{BF1,Munich,Q2evo,HT}
%\cite{BF1}-\cite{HT},
%On the other hand, 
various groups have been able to fit
the available data 
%({\it separately at low and high $Q^2$ values})
using a hard input at small $x$: 
$x^{-\lambda},~\lambda >0$ with different $\lambda$ values at low and high 
$Q^2$ (see \cite{LoYn}-\cite{DeJePa}).
%\cite{LoYn,DoLa,Abramo,YF93,FKR,CaKaMeTTV}).
Such results are well-known at low $Q^2$ values \cite{DoLa}. 
At large $Q^2$ values, for 
%In some sense, it is not very surprising, because 
the modern HERA data 
it is also not very surprising, because they
cannot distinguish between the behavior
based on a steep input parton parameterization,
at quite large $Q^2$, and the
steep form acquired after the dynamical evolution from a flat initial
condition at quite low $Q^2$ values.

As it has been mentioned above and shown in \cite{Q2evo,HT},
the behavior of parton densities and $F_2$ given in the Bessel-like form 
by generalized DAS approach
%Eqs. (\ref{9.10})-(\ref{9}) 
can mimic a power law shape
over a limited region of $x$ and $Q^2$
 \begin{eqnarray}
f_a(x,Q^2) \sim x^{-\lambda^{\rm eff}_a(x,Q^2)}
 ~\mbox{ and }~
F_2(x,Q^2) \sim x^{-\lambda^{\rm eff}_{\rm F_2}(x,Q^2)}.
\nonumber    \end{eqnarray}

The effective slopes $\lambda^{\rm eff}_a(x,Q^2)$ and $\lambda^{\rm eff}_{\rm F_2}(x,Q^2)$
have the form:
 \begin{eqnarray}
\lambda^{\rm eff}_g(x,Q^2) &=& \frac{f^+_g(x,Q^2)}{f_g(x,Q^2)} \,
\rho \, \frac{\tilde I_1(\sigma)}{\tilde I_0(\sigma)},
\nonumber
\\
\lambda^{\rm eff}_q(x,Q^2) &=& \frac{f^+_q(x,Q^2)}{f_q(x,Q^2)} \,
\rho \, \frac{\tilde  I_2(\sigma) (1- 20 a_s(Q^2))
 + 20 a_s(Q^2) \tilde I_1(\sigma)/\rho}{\tilde  I_1(\sigma) 
(1- 20 a_s(Q^2))
 + 20 a_s(Q^2) \tilde I_0(\sigma)/\rho},
\nonumber
%\label{10.1}
%
\\
\lambda^{\rm eff}_{\rm F_2}(x,Q^2) &=& \frac{\lambda^{eff}_q(x,Q^2) \,
f^+_q(x,Q^2) + (2f)/3a_s(Q^2)\, \lambda^{eff}_g(x,Q^2) \,
f^+_g(x,Q^2)}{f_q(x,Q^2) + (2f)/3 a_s(Q^2)\, f_g(x,Q^2)},
%\nonumber
\label{10.1}
\end{eqnarray}
where 
the exact form of parton densities can be found in \cite{Q2evo,HT}.

The results (\ref{10.1}) (and also (\ref{11.1a})--(\ref{11.1}) below) are 
given at the
NLO approximation. To obtain the LO
one, it is necessary to cancel the term $\sim a_s(Q^2)$ and to use Eqs. 
(\ref{8.0})--(\ref{8.02}) for parton densities $f_a(x,Q^2)$.

The effective slopes $\lambda^{\rm eff}_a $ and 
$\lambda^{\rm eff}_{\rm F_2}$ depend on the magnitudes $A_a$ of the initial PDFs
and also on the chosen input values of $Q^2_0$ and $\Lambda $.
To compare with the experimental data it is necessary the exact expressions
(\ref{10.1}), but for qualitative analysis it is better to use an 
approximation.

\subsection{Asymptotic form of the effective slopes}

At quite 
large values of $Q^2$, 
where the ``$-$'' component is negligible,
the dependence on the initial PD disappears, having
in this case for the asymptotic behavior the following 
expressions\footnote{The asymptotic formulae given in 
Eqs. (\ref{11.1a})--(\ref{11.1})
work quite well at any $Q^2 \geq Q^2_0$ values,
because at $Q^2=Q^2_0$ the values of
$\lambda^{\rm eff}_a $ and $\lambda^{\rm eff}_{\rm F_2}$ are equal zero. 
The use of approximations in Eqs. (\ref{11.1a})--(\ref{11.1}) instead of the exact results 
given in Eq. (\ref{10.1}) underestimates 
(overestimates) only slightly the gluon (quark) slope
at $Q^2 \geq Q^2_0$.
%For the $F_2$ case, the similarity of $\lambda^{eff}_{F2} $ and 
%$\lambda^{eff,as}_{F2}$ values is shown in Fig 1.
}:
 \begin{eqnarray}
\lambda^{\rm eff,as}_g(x,Q^2) &=& 
\rho\, \frac{\tilde I_1(\sigma)}{\tilde I_0(\sigma)} \approx \rho - 
\frac{1}{4\ln{(1/x)}}, 
%\nonumber \\
\label{11.1a} \\
\lambda^{\rm eff,as}_q(x,Q^2) &=& 
\rho \frac{\tilde  I_2(\sigma) (1- 20 a_s(Q^2))
 + 20 a_s(Q^2)\tilde  I_1(\sigma)/\rho}{\tilde  I_1(\sigma) 
(1- 20 a_s(Q^2))
 + 20 a_s(Q^2)\tilde  I_0(\sigma)/\rho}
 \nonumber \\
%&=& \rho\, \frac{\tilde I_2(\sigma)}{\tilde I_1(\sigma)} 
% + 20 a_s(Q^2)\left( 1- \frac{\tilde  I_0(\sigma) 
%\tilde  I_2(\sigma)}{\tilde  I_1^2(\sigma)} \right)
&\approx &
 \rho - \frac{3}{4\ln{(1/x)}} +  
\frac{10a_s(Q^2)}{ \rho \ln{(1/x)}}, 
%
% \nonumber \\
\label{11.1b}\\ 
\lambda^{\rm eff,as}_{\rm F_2}(x,Q^2) 
&=& \rho\, \frac{\tilde I_2(\sigma)}{\tilde I_1(\sigma)} 
 + 26 a_s(Q^2)\left( 1- \frac{\tilde  I_0(\sigma) 
\tilde  I_2(\sigma)}{\tilde  I_1^2(\sigma)} \right)
 \nonumber \\
%&=& \lambda^{\rm eff,as}_q(x,Q^2) + 6 a_s(Q^2)\left( 1- \frac{\tilde  I_0(\sigma) 
%\tilde  I_2(\sigma)}{\tilde  I_1^2(\sigma)} \right)
% \nonumber \\
&\approx &
 \rho - \frac{3}{4\ln{(1/x)}} +  
\frac{13a_s(Q^2)}{ \rho \ln{(1/x)}} 
= \lambda^{\rm eff,as}_q(x,Q^2)  +  
\frac{3a_s(Q^2)}{ \rho \ln{(1/x)}},
%\nonumber
\label{11.1} 
\end{eqnarray}
where the symbol $\approx $ marks the approximation obtained in the  expansion
of the usual and modified Bessel functions in (\ref{4}).
% $I_n(\sigma)$ $(n=0,1,2)$. 
These approximations are
accurate only at very large $\sigma $ values (i.e. at very large $Q^2$
and/or very small $x$).

As one can see from Eqs. (\ref{11.1a}) and (\ref{11.1b}),  
%As it has been already shown in Ref. \cite{Q2evo},
the gluon effective slope $\lambda^{\rm eff}_g$ 
is larger than the quark slope
$\lambda^{\rm eff}_q$, which is in excellent agreement with 
%a recent
MRS \cite{MRS} and GRV \cite{GRVold} analyses.
% (see also \cite{fits}). 

We would like to note that at the NLO approximation the slope 
$\lambda^{\rm eff,as}_{\rm F_2}(x,Q^2)$ lies between quark and
gluon ones but closely to quark slope 
$\lambda^{\rm eff,as}_{q}(x,Q^2)$.
Indeed,
 \begin{eqnarray}
 \lambda^{\rm eff,as}_{g}(x, Q^2) \ - \
 \lambda^{\rm eff,as}_{\rm F_2}(x, Q^2) \ 
%&=& \
% \left(\rho \, \frac{\widetilde{I}_1(\sigma)}{\widetilde{I}_0(\sigma)}
% + 26 a_s(Q^2) \right)
%\left(1 -
%   \frac{\widetilde{I}_0(\sigma) \widetilde{I}_2(\sigma)}
%         {\widetilde{I}_1^2(\sigma)} \right)
%\nonumber \\
 \ &\approx& \ \left(\rho - \frac{1}{4\ln{(1/x)}}
 + 26 a_s(Q^2)\right) \frac{1}{2 \rho \ln{(1/x)}}  ,
\label{Slopes:NLO:G-F2} \\
 \lambda^{\rm eff,as}_{\rm F_2}(x, Q^2) \ - \
 \lambda^{\rm eff,as}_{q}(x, Q^2) \ 
%&=& \
% 6 \, a_s(Q^2) \left(1 -
%   \frac{\widetilde{I}_0(\sigma) \widetilde{I}_2(\sigma)}
 %        {\widetilde{I}_1^2(\sigma)} \right) \ \
&\approx &\
 \frac{3 a_s(Q^2)}{\rho \ln(1/x)}  .
\label{Slopes:NLO:F2-q}
 \end{eqnarray}

Both slopes $\lambda^{\rm eff}_a(x, Q^2)$ decrease with increasing $x$ (see Fig. 5). \
A $x$-dependence of the slope should not appear for PDFs within a
Regge type asymptotic ($x^{-\lambda}$) and precise measurement of the slope 
$\lambda^{\rm eff}_a(x, Q^2)$ may lead to the possibility to verify the
type of small-$x$ asymptotics of parton distributions.

\section{$F_2^{cc}$ and  $F_2^{bb}$ structure functions} \indent

%{\bf Heavy quark part}\\
%Last year
Recently the H1 \cite{Collaboration:2009jy,:2009ut} and ZEUS
\cite{Abramowicz:2010zq,Chekanov:2009kj}
Collaborations at HERA presented new data\footnote{The papers
\cite{Collaboration:2009jy}-\cite{Chekanov:2009kj}
%\cite{Collaboration:2009jy,:2009ut,Abramowicz:2010zq,Chekanov:2009kj}
contain also the references on the previous data on
deep-inelastic (DIS)
structure functions (SFs) $F_2^{cc}$ and $F_2^{bb}$ at
 small $x$ values.}.
Moreover, the preliminary combine H1 and ZEUS data of $F_2^{cc}(x,Q^2)$ and
$F_2^{bb}(x,Q^2)$
has been demonstrated
%shown
recently (see \cite{Lipka:2009zza}).

In the framework of DGLAP
%Dokshitzer-Gribov-Lipatov-Altarelli-Parisi (DGLAP)
dynamics \cite{DGLAP}, there are two basic methods to study
heavy-flavour physics.
One of them \cite{Kniehl:1996we} is based on the massless PDF evolution of parton
%distribution functions (PDFs) 
and the other one on the photon-gluon fusion (PGF) process
\cite{Frixione:1994dv}.
There are also some interpolating schemes (see
Ref.~\cite{Olness:1987ep} and references cited therein).
%The present HERA data on $F_2^c$
%\cite{Aktas:2004az,Aktas:2005iw,Chekanov:2003rb,Chekanov:2007ch} are in good
%agreement with the modern theoretical predictions.

%In this short paper,
%%in the framework of PGF 
Here we present the results of Ref. \cite{Illarionov:2011km} were we
applied compact low-$x$
approximation formulae for the
SFs $F_2^{ii}(x,Q^2)$, with hereafter $i=c,b$, observed \cite{Illarionov:2008be}
in the framework of PGF process at the first two orders
of perturbation theory to these new HERA experimental data
\cite{Collaboration:2009jy}-\cite{Lipka:2009zza}.
%\cite{Collaboration:2009jy,:2009ut,Abramowicz:2010zq,Chekanov:2009kj,Lipka:2009zza}.
%at leading order (LO) and  next-to-leading order (NLO), which lead to
%become to be in
We show a good agreement between experimental data and the approach which found
%with modern experimental data
without additional  free parameters. All
%set of
PDF parameters
%of parton distribution functions (PDFs)
have been fitted
earlier \cite{HT,Cvetic:2009kw} from  $F_2(x,Q^2)$ HERA
experimental data.

In the framework of the generalized DAS approach, the SFs  $F_2^{cc}(x,Q^2)$ and  $F_2^{bb}(x,Q^2)$ 
have the following form
%%{\bf \large Wilson coefficients}\\
%One has $M_{2,g}^+(1)=M_{2,g}^-(1)$ if
%$M_{2,g}^r(n)$ are devoid of singularities in the limit $\delta_r\to0$, as
%we assume for the time being.
%Such singularities actually occur at NLO, leading to modifications to be
%discussed below.
%Defining $M_{2,g}(1)=M_{2,g}^\pm(1)$ and using
%$f_g(x,Q^2)=\sum_{r=\pm}f_g^r(x,Q^2)$, Eq.~(\ref{eq:pm1}) may be simplified to
%become
\begin{equation}
F_2^{ii}(x,Q^2) \approx M^i_{2,g}(1,Q^2,\mu^2) f_g(x,\mu^2), ~~ (i=c,b)
\label{eq:pm2n}
\end{equation}
where 
%usually $\mu^2=Q^2+4m^2_i$ is used. Here 
$M^i_{2,g}(1,Q^2,\mu^2)$ is the first Mellin moment of the so-called gluon 
coefficient function $C^i_{2,g}(x,Q^2,\mu^2)$.

Through NLO, $M_{2,g}(1,Q^2,\mu^2)$ exhibits the structure
%\begin{eqnarray}
\be
M^i_{2,g}(1,Q^2,\mu^2) = e_i^2a_s(\mu)\left\{M_{2,g}^{(0)}(1,c_i)
+a_s(\mu)\left[M_{2,g}^{(1)}(1,c_i)+M_{2,g}^{(2)}(1,c_i) \ln\frac{\mu^2}{m_i^2}
\right]\right\}+{\mathcal O}(a_s^3),
\label{eq:exp}
%\end{eqnarray}
\ee
where 
\be c_i=\frac{m_i^2}{Q^2},~~~\mu^2=Q^2+4m^2_i \, .  \label{c_i}. \ee

%\vskip 0.5cm

\subsection{LO results}
\label{sec:lo}

%{\bf LO}\\
The LO coefficient function of PGF can be obtained from the QED case
\cite{Baier:1966bf} by adjusting coupling constants and colour factors, and
they read \cite{Witten:1975bh,Kotikov:2001ct}
%\begin{eqnarray}
\be
C_{2,g}^{(0)}(x,c) ~=~ -2x\{[1-4x(2-c)(1-x)]\beta
-[1-2x(1-2c)
+2x^2(1-6c-4c^2)]L(\beta)\},
\label{eq:lo}
%\end{eqnarray}
\ee
where
\begin{equation}
\beta(x)=\sqrt{1-\frac{4cx}{1-x}},\qquad
L(\beta)=\ln\frac{1+\beta}{1-\beta}.
\label{eq:lo1}
\end{equation}

%We perform
Performing
the Mellin transformation 
%\cite{Illarionov:2008be} in Eq.~(\ref{eq:mel}),
\be
M_{2,g}(n,c) ~=~ \int^b_0 \, \frac{dx}{x} \, C_{2,g}(x,c)
\label{Mellin}
\end{equation}
we find at $n=1$ (see \cite{Illarionov:2008be})
\footnote{Note that similar formulas work well for (see 
\cite{Illarionov:2011wc}) for  high-energy neutrino-nucleo scattering
where the effective value of the Bjorken variable $x$ is very small.}
\be
M_{2,g}^{(0)}(1,c) ~=~ \frac{2}{3}[1+2(1-c)J(c)]
\label{lo2}
\ee
with
\be
J(c) = - \sqrt{b}\ln t,\qquad t=\frac{1-\sqrt{b}}{1+\sqrt{b}},\qquad
b=\frac{1}{1+4c}.
\label{eq:lo2a}
\ee

%\vskip 0.5cm
\subsection{NLO results}
\label{sec:nlo}

%{\bf NLO}\\
The NLO coefficient functions of PGF are rather lengthy and not published in
print; they are only available as computer codes \cite{Laenen:1992zk}.
For the purpose of this letter, it is sufficient to work in the high-energy
regime, defined by $x\ll1$, where they assume the compact form
\cite{Catani:1992zc}
\begin{equation}
C_{2,g}^{(j)}(x,c)=\beta R_{2,g}^{(j)}(1,c),
\label{eq:nlo}
\end{equation}
with
\be
R_{2,g}^{(1)}(1,c)~=~\frac{8}{9}C_A[5+(13-10c)J(c)+6(1-c)I(c)],\qquad
R_{2,g}^{(2)}(1,c)~=~-4 C_A M_{2,g}^{(0)}(1,c),
\label{eq:nloA}
\ee
where $C_A=N$ for the colour gauge group SU(N), $J(c)$ is defined by
Eq.~(\ref{eq:lo2a}), and
\begin{equation}
I(c)=-\sqrt{b}\left[\zeta(2)+\frac{1}{2}\ln^2t-\ln(bc)\ln t+2\Li_2(-t)\right],
\label{eq:nlo1}
\end{equation}
where $t$ is given in (\ref{eq:lo2a}) and
%$\zeta(2)=\pi^2/6$ and
$\Li_2(x)=-\int_0^1(dy/y)\ln(1-xy)$ is the dilogarithmic function.

As already mentioned above (see the end of Section 2), the Mellin transforms of
$C_{k,g}^{(j)}(x,c)$ exhibit singularities in the limit $\delta_{\pm}\to0$, which
lead to modifications in 
%Eqs.~(\ref{eq:pm1}) and 
Eq. (\ref{eq:pm2n}).
As was shown in Refs.~\cite{YF93,Q2evo,HT},
the terms involving $1/\delta_{\pm}$ correspond to singularities of 
the Mellin moments $M_{2,g}^{\pm}(n)$ at $n \to 1$ and
depend on the exact form of the subasymptotic
low-$x$ behaviour encoded in $\tilde{f}_g^{\pm}(x,\mu^2)$.
%, as
The modification is simple:
\begin{equation}
\frac{1}{\delta_{\pm}} \to \frac{1}{\tilde \delta_{\pm}},\qquad
\frac{1}{\tilde \delta_{\pm}}=\frac{1}{\tilde{f}_g^{\pm}(\hat{x},\mu^2)}
\int^1_{\hat{x}}\frac{dy}{y}\tilde{f}_g^{\pm}(y,\mu^2),
\label{eq:nlo2}
\end{equation}
where $\hat{x}=x/b$.
In the generalized DAS
%double-asymptotic scaling 
regime, the $+$ and $-$
components of the gluon PDF exhibit the
%following
low-$x$ behaviour (\ref{8.0A})-(\ref{8.02b}).
We thus have \cite{Q2evo,HT}
\begin{equation}
\frac{1}{\tilde \delta_+}
%=
\approx \frac{1}{\rho(\hat{x})}\,
\frac{I_1(\sigma(\hat{x}))}{I_0(\sigma(\hat{x}))},
\qquad
\frac{1}{\tilde \delta_-}
%=
\approx \ln\frac{1}{\hat{x}},
%\approx\frac{1}{\hat{\rho}}-\frac{\beta_0}{48s},
\label{eq:nlo3}
\end{equation}
where $\sigma$ and $\rho$ are given in (\ref{intro:1}).

%where $\hat{\sigma}$ and $\hat{\rho}$ are $\sigma$ and $\rho$ evaluated at
%$x=\hat{x}$, respectively.
Because the ratio $f_g^-(x,Q^2)/f_g^+(x,Q^2)$ is rather small at the $Q^2$
values considered, Eq.~(\ref{eq:pm2n}) is modified to become
\begin{equation}
F_2^{ii}(x,Q^2)\approx\tilde{M}_{2,g}(1,\mu^2,c_i) f_g(x,\mu^2),
\end{equation}
where $\tilde{M}_{2,g}(1,\mu^2)$ is obtained from $M_{2,g}(n,\mu^2)$ by taking
the limit $n\to 1$ and replacing $1/(n-1)\to1/\tilde \delta_+$.
Consequently, one needs to substitute
\begin{equation}
M_{2,g}^{(j)}(1,c)\to\tilde{M}_{2,g}^{(j)}(1,c)\quad(j=1,2)
\end{equation}
in the NLO part of Eq.~(\ref{eq:exp}).
Using the identity
\begin{equation}
\frac{1}{I_0(\sigma(\hat{x}))}
\int^1_{\hat{x}}\frac{dy}{y}\beta\left(\frac{x}{y}\right) I_0(\sigma(y))
%=
\approx \frac{1}{\tilde \delta_+}-\ln (bc)-\frac{J(c)}{b},
\end{equation}
we find the Mellin transform of Eq.~(\ref{eq:nlo}) to be
\footnote{Note, that $\delta_+$ determines the behavior of the slope of
gluon density (see (\ref{10.1}))
and also mostly the  slope of SF $F_2$. The form (\ref{eq:nlo3})
of $\tilde \delta_+$ is in full agreement 
%\cite{Kotikov:2002fd}
with the results (\ref{10.1}) for the asymptotic form of the effective slope of gluon density.}
%corresponding HERA experimentaldata.}
\begin{equation}
\tilde{M}_{2,g}^{(j)}(1,c)\approx
%=
\left[\frac{1}{\delta_+}-\ln(bc)-\frac{J(c)}{b}\right]
R_{2,g}^{(j)}(1,c)\quad(j=1,2),
\label{nloA}
\end{equation}
with $R_{2,g}^{(j)}(1,a)\quad(j=1,2)$ are given in (\ref{eq:nloA}).
The rise of the NLO terms as $x\to0$ is in agreement with earlier
investigations \cite{Nason:1987xz}.

\begin{figure}[t]
\includegraphics[height=0.55\textheight,width=0.95\textwidth]{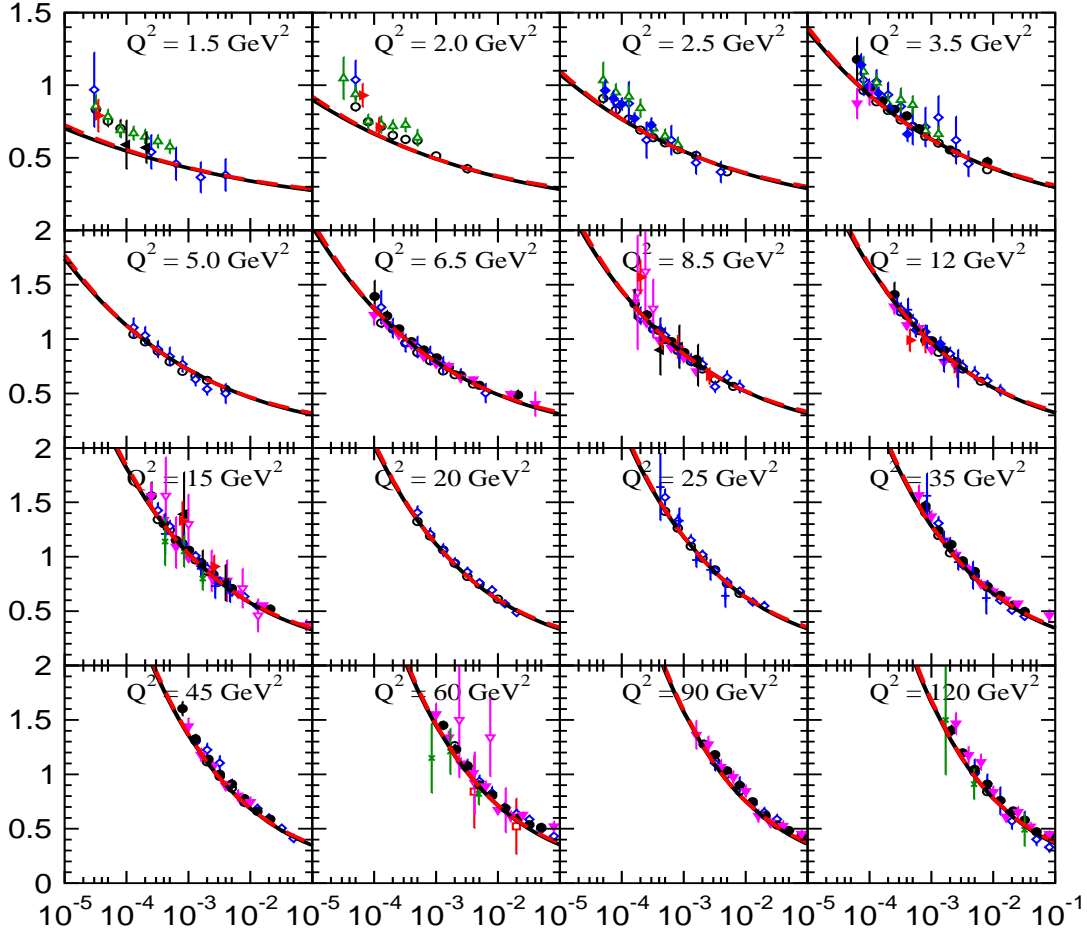}
\vskip 1.3cm
\caption{$F_2(x,Q^2)$ as a function of $x$ for different $Q^2$ bins. 
The experimental points are from H1 \cite{H197} (open points) and ZEUS 
\cite{ZEUS01} (solid points) at 
$Q^2 \geq 1.5$ GeV$^2$.
The solid curve represents the NLO fit. The dashed curve (hardly 
distinguishable 
from the solid one) represents the LO fit.}
\label{fig1}
\end{figure}

\section{Comparison with experimental data for SF $F_2$ and the slope $\lambda_{\rm F_2}$
%Results of the fits
} \indent

Using the results of previous section we have
analyzed  HERA data for $F_2$ and the slope $\partial \ln F_2/\partial \ln (1/x)$
at small $x$ from the H1 and ZEUS Collaborations \cite{H197}-\cite{DIS02}.
%\cite{H197,ZEUS01,Surrow,H1slo,DIS02}.

In order to keep the analysis as simple as possible,
we fix $f=4$ and $\alpha_s(M^2_Z)=0.1166 $ (i.e., $\Lambda^{(4)} = 284$ MeV) in agreement
with the more recent ZEUS results \cite{ZEUS01}.

\begin{figure}[t]
\includegraphics[height=0.55\textheight,width=0.95\textwidth]{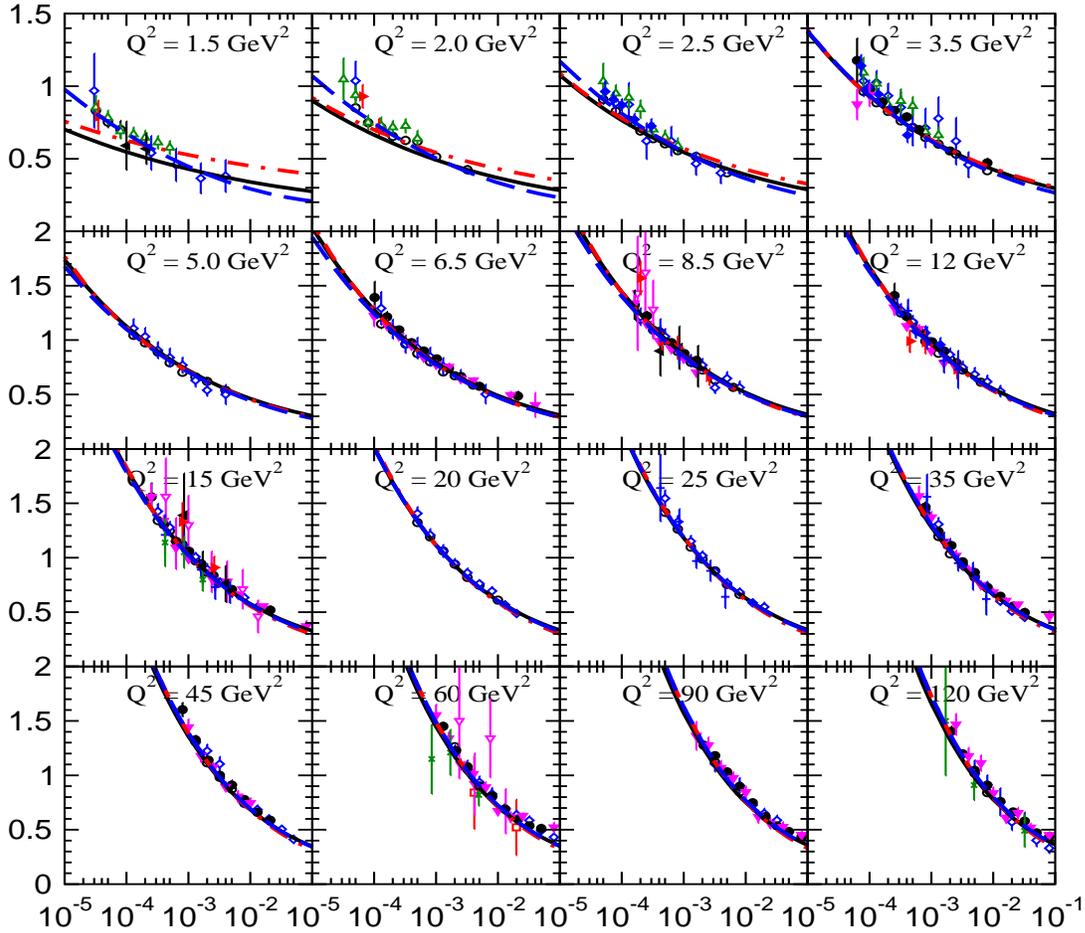}
\vskip 1.3cm
\caption{$F_2(x,Q^2)$ as a function of $x$ for different $Q^2$ bins. 
The experimental points are same as on Fig. 1.
The solid curve represents the NLO fit. 
The dash-dotted curve represents the BFKL-motivated estimation for higher-twist
corrections to $F_2(x,Q^2)$ (see \cite{HT}).
The dashed curve is obtained from the fits at the NLO, when the renormalon contributions
of higher-twist terms have been incorporated.}
\label{fig2}
\end{figure}

\begin{figure}[t]
\includegraphics[height=0.55\textheight,width=0.95\textwidth]{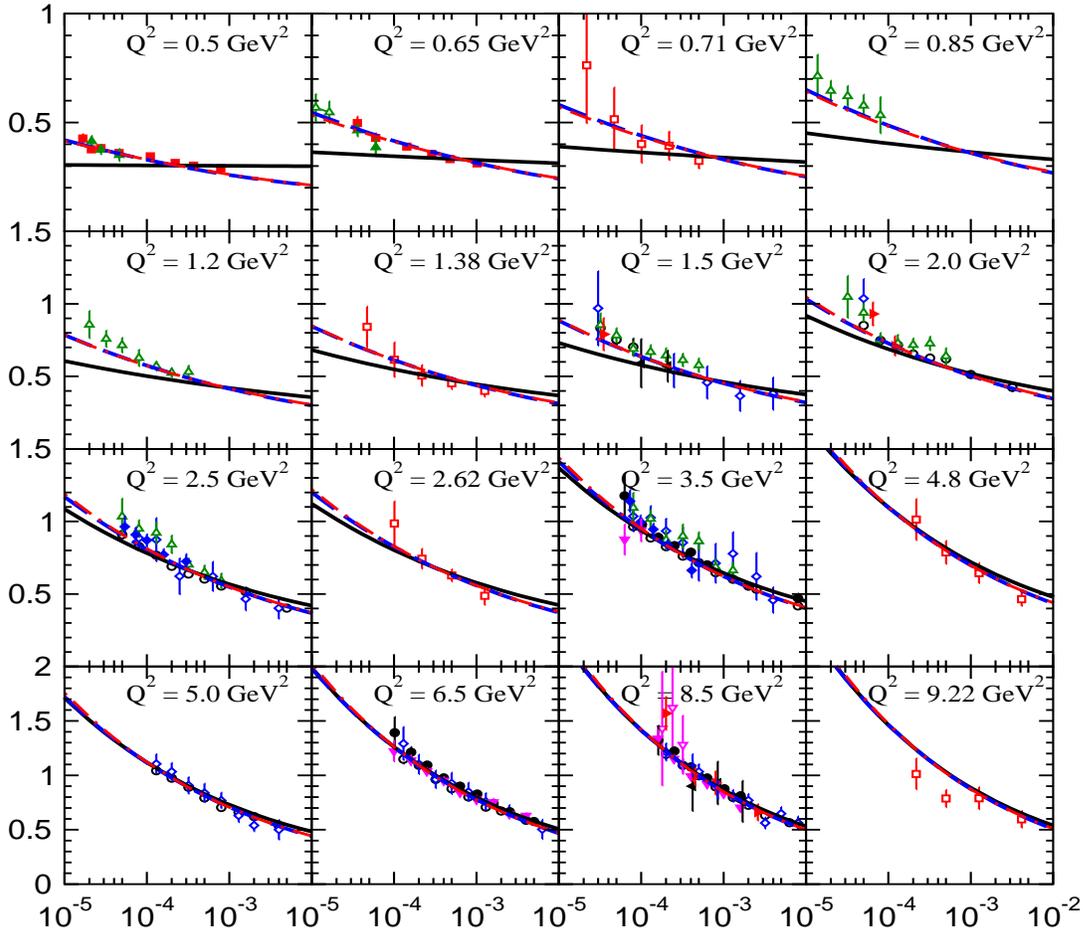}
\vskip 1.5cm
\caption{$F_2(x,Q^2)$ as a function of $x$ for different $Q^2$ bins. 
The experimental points are from H1 \cite{H197} 
  (open points) and ZEUS \cite{ZEUS01} (solid points)
at $Q^2 \geq 0.5$ GeV$^2$.
The solid curve represents the NLO fit. 
The dashed curve is from the fits at the NLO with the renormalon contributions
of higher-twist terms incorporated.
The dash-dotted curve (hardly distinguishable 
from the dashed one) represents the LO fit with the renormalon contributions
of higher-twist terms incorporated.}
\label{fig3}
\end{figure}

\begin{figure}[t]
\includegraphics[height=0.55\textheight,width=0.95\textwidth]{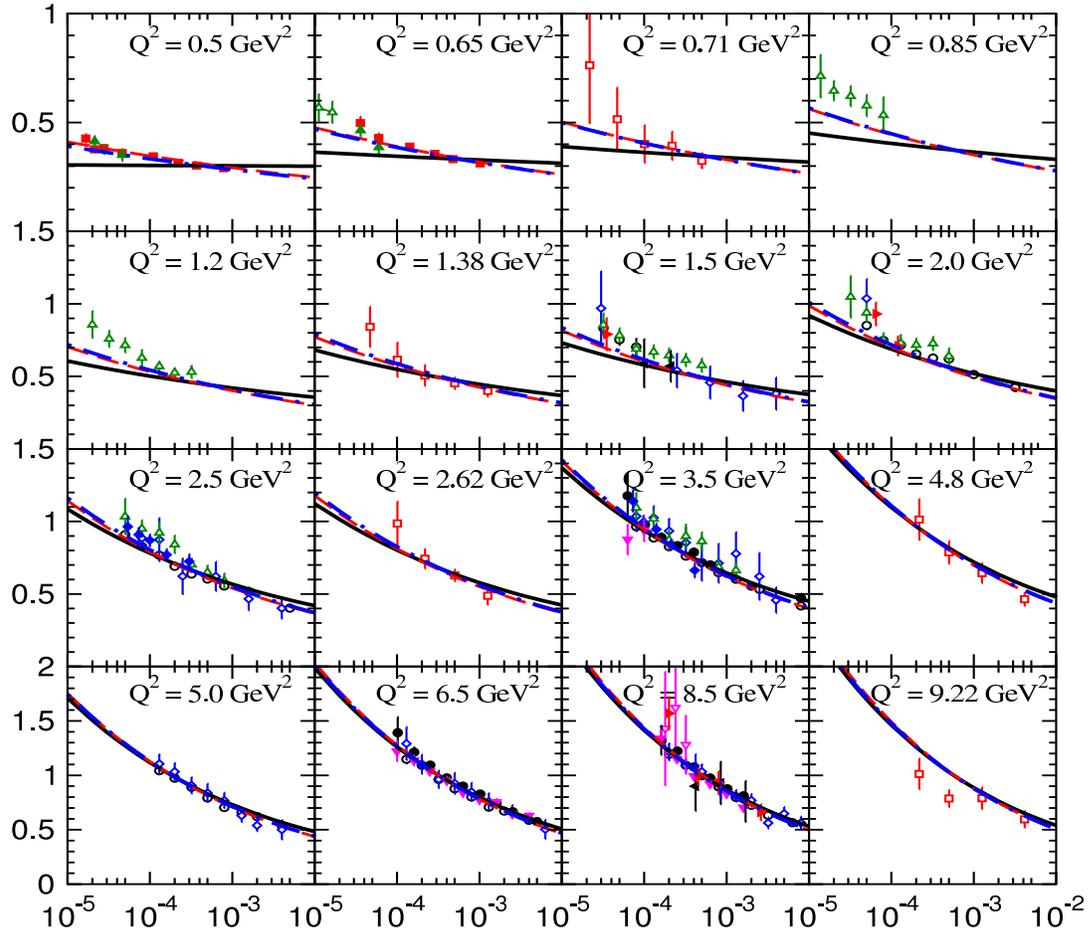}
\vskip 1.5cm
\caption{ $x$ dependence of $F_2(x, Q^2)$ in bins of $Q^2$. The experimental data from H1 (open points) 
and ZEUS (solid points) are compared with the NLO fits for $Q^2\geq 0.5$ GeV$^2$
implemented with the canonical (solid lines), frozen (dot-dashed lines), and analytic (dashed lines) 
versions of the strong-coupling constant. For comparison, also the results
obtained in Ref. \cite{HT} through a fit based on the renormalon model of higher-twist terms are shown 
(dotted lines).}
\label{fig3}
\end{figure}

%\begin{\Large}
\begin{table}
\caption{
%\label{Tab:H1+ZEUS:96/97}\sffamily
The result of the LO and NLO fits to H1 
%(1996/97) \protect\cite{Adloff:1999}
and ZEUS 
%(1996/97) \protect\cite{Chekanov:2001} 
data  for different low
$Q^2$ cuts.  In the fits $f$ is fixed to 4 flavors.
}
\centering
\footnotesize
%\small
\large
\vspace{0.3cm}
%\begin{ruledtabular}
\begin{tabular}{|l||c|c|c||r|} \hline \hline
& $A_g$ & $A_q$ & $Q_0^2~[{\rm GeV}^2]$ &
 $\chi^2 / n.o.p.$~ \\
\hline\hline
~$Q^2 \geq 1.5 {\rm GeV}^2 $  &&&& \\
 LO & 0.784$\pm$.016 & 0.801$\pm$.019 & 0.304$\pm$.003 & 754/609 \\
 LO$\&$an. & 0.932$\pm$.017 & 0.707$\pm$.020 & 0.339$\pm$.003 & 632/609  \\
  LO$\&$fr. & 1.022$\pm$.018 & 0.650$\pm$.020 & 0.356$\pm$.003 & 547/609   \\
\hline
 NLO & -0.200$\pm$.011 & 0.903$\pm$.021 & 0.495$\pm$.006 & 798/609 \\
 NLO$\&$an. & 0.310$\pm$.013 & 0.640$\pm$.022 & 0.702$\pm$.008 & 655/609  \\
  NLO$\&$fr. & 0.180$\pm$.012 & 0.780$\pm$.022 & 0.661$\pm$.007 & 669/609   \\
\hline\hline
~$Q^2 \geq 0.5 {\rm GeV}^2 $  &&&& \\
 LO & 0.641$\pm$.010 & 0.937$\pm$.012 & 0.295$\pm$.003 & 1090/662 \\
 LO$\&$an. & 0.846$\pm$.010 & 0.771$\pm$.013 & 0.328$\pm$.003 & 803/662  \\
  LO$\&$fr. & 1.127$\pm$.011 & 0.534$\pm$.015 & 0.358$\pm$.003 & 679/662   \\
\hline
 NLO & -0.192$\pm$.006 & 1.087$\pm$.012 & 0.478$\pm$.006 & 
{\color{red} 1229/662} \\
 NLO$\&$an. & 0.281$\pm$.008 & 0.634$\pm$.016 & 0.680$\pm$.007 & 
{\color{red} 633/662}  \\
  NLO$\&$fr. & 0.205$\pm$.007 & 0.650$\pm$.016 & 0.589$\pm$.006 & 
{\color{red} 670/662}   \\
\hline \hline
%\normale
\end{tabular}
%\end{ruledtabular}
\end{table}

\begin{figure}[htb]
\begin{center}
\includegraphics[height=4.5in,width=5.8in]{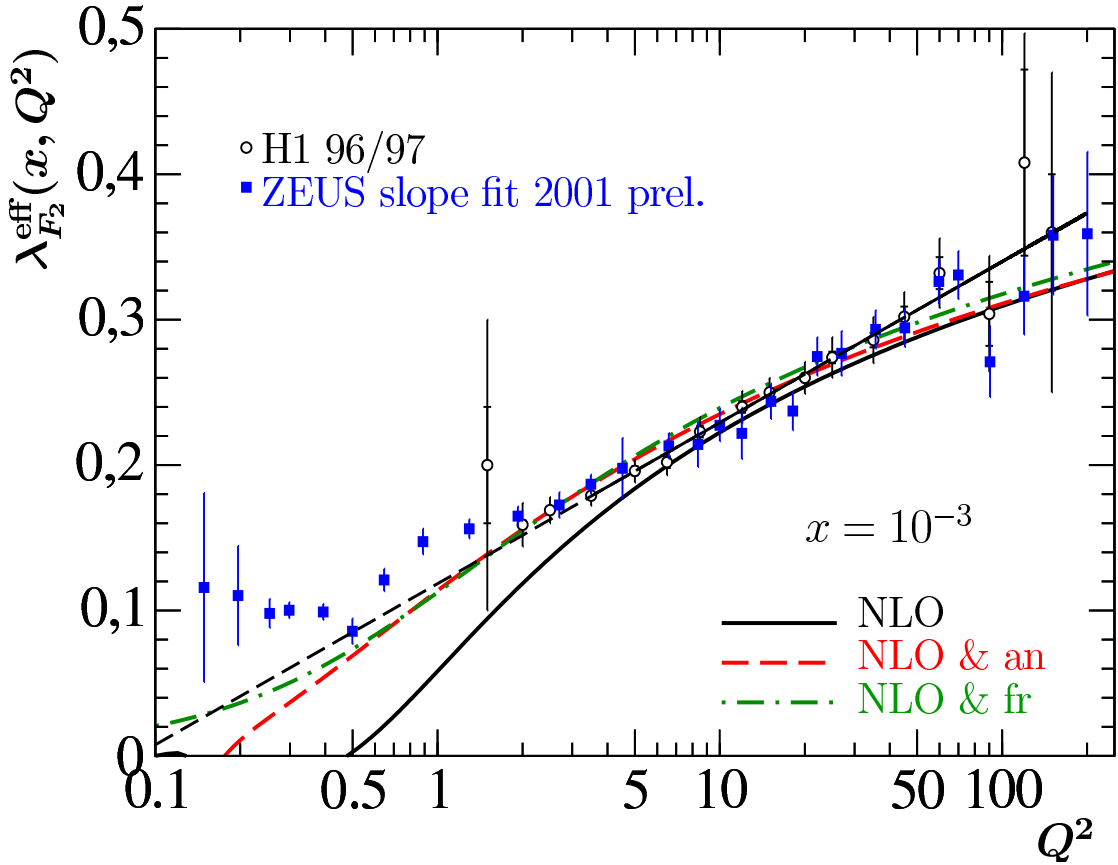}
\end{center}
%\vskip 0.5cm
\caption{The values of effective slope
$\lambda^{\rm eff}_{\rm F_2}$ 
as a function of $Q^2$ for $x=10^{-3}$.  
The experimental points are from H1 \cite{H1slo,DIS02} (open points) and ZEUS 
\cite{Surrow} (solid points). The solid curve represents the NLO fit.
The dash-dotted  and lower dashed curves represent the NLO fits with 
``frozen'' and analytic coupling constants, respectively. The top dashed line
 represents the fit from \cite{H1slo}.}
\label{fig4}
\end{figure}

\begin{figure}[htb]
\begin{center}
\includegraphics[height=4.5in,width=5.8in]{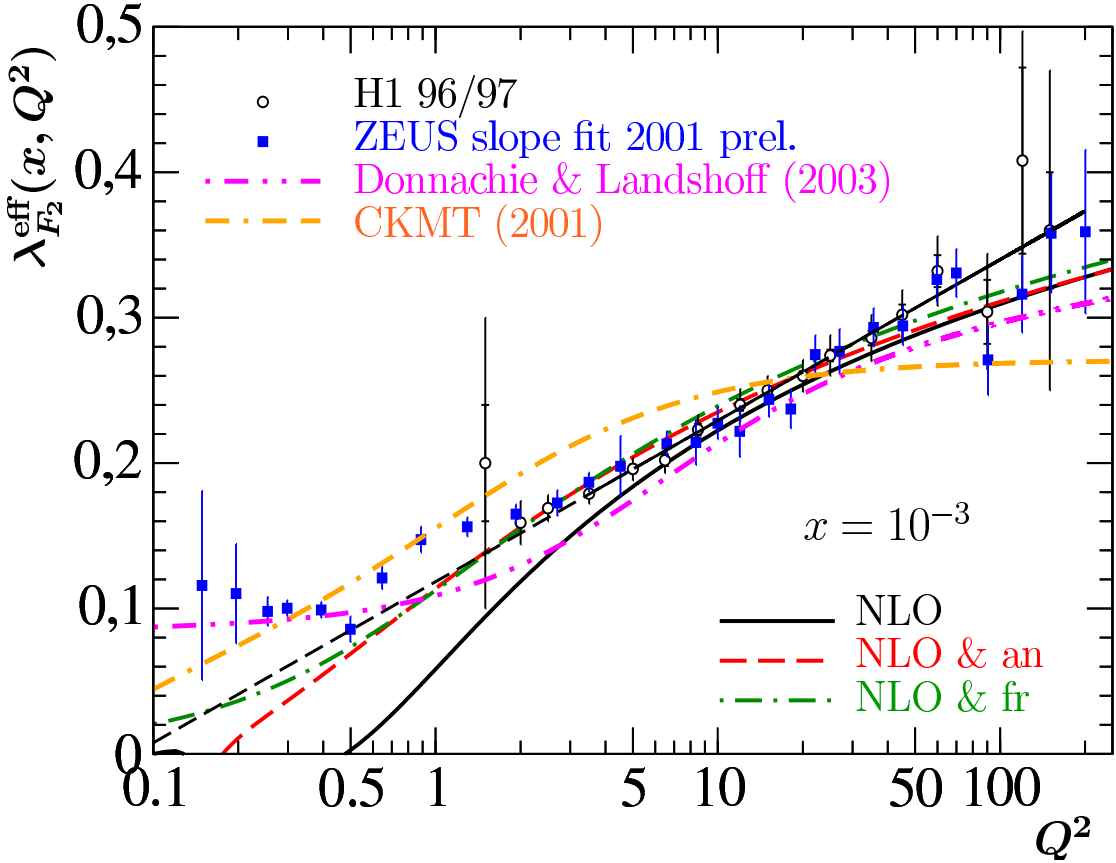}
\end{center}
%\vskip 1cm
\caption{
$Q^2$ dependence of $\lambda^{\rm eff}_{\rm F_2}(x,Q^2)$
for an average small-$x$ value of $x = 10^{-3}$. The
experimental data from H1 (open points) and ZEUS (solid points) are compared
with the NLO fits for $Q^2 \geq 0.5$ GeV$^2$ implemented with the canonical (solid line),
frozen (dot-dashed line), and analytic (dashed line) versions of the strong-coupling
constant. The linear rise of $\lambda^{\rm eff}_{\rm F_2}(x,Q^2)$
 with $\ln Q^2$ as described by Eq. (2) is indicated
by the straight dashed line. For comparison, also the results obtained in the
phenomenological models by Capella et al.
%Kaidalov et al. 
%[47] 
\cite{CaKaMeTTV} (dash-dash-dotted line) and by
Donnachie and Landshoff \cite{Donnachie:2003cs} 
(dot-dot-dashed line) are shown.}
\label{fig5}
\end{figure}

\begin{figure}[htb]
\begin{center}
\includegraphics[height=4.5in,width=5.8in]{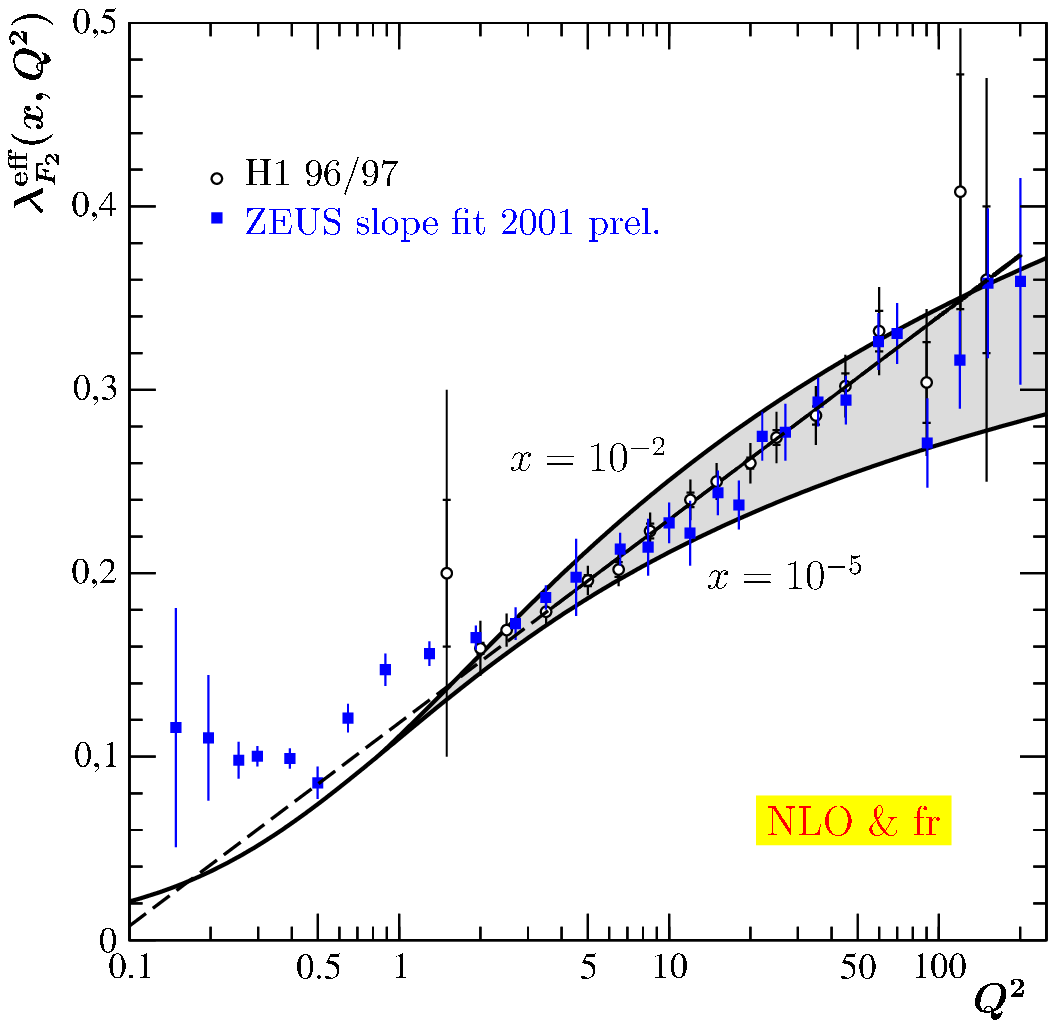}
\end{center}
%\vskip 1cm
\caption{The values of effective slope
$\lambda^{\rm eff}_{\rm F_2}$  
as a function of $Q^2$.
The experimental points are same as on Fig. 4.
The dashed line
 represents the fit from \cite{H1slo}.
The solid curves represent the NLO fits with 
``frozen'' coupling constant
at $x=10^{-2}$ and  $x=10^{-5}$.}
\label{fig5}
\end{figure}

As it is possible to see in Fig. 1 (see also \cite{Q2evo,HT}), the twist-two
approximation is reasonable at $Q^2 \geq 2$ GeV$^2$. At smaller $Q^2$, some
modification of the approximation should be considered. In Ref.
%the recent article
\cite{HT} we have added the higher twist corrections.
For renormalon model of higher twists, we
have found a good
agreement with experimental data at essentially lower $Q^2$ values:
$Q^2 \geq 0.5$ GeV$^2$ (see Figs. 2 and 3), but we have added 4 additional parameters:
amplitudes of twist-4 and twist-6 corrections to quark and gluon densities.

Moreover, 
 the results of fits in \cite{HT} have an important property: they are
very similar in LO and NLO approximations of perturbation theory.
The similarity is related to the fact that the small-$x$ asymptotics of 
the NLO corrections
are usually large and negative (see, for example, $\alpha_s$-corrections 
\cite{FaLi,KoLi} to
BFKL kernel \cite{BFKL}\footnote{It seems that it is a property of 
any processes in which gluons,
but not quarks play a basic role.}).
% and 
Then, the LO form $\sim \alpha_s(Q^2)$ for
some observable and the NLO one 
$\sim \alpha_s(Q^2) (1-K\alpha_s(Q^2)) $
with a large value of $K$ are similar, because 
%usually 
$\Lambda \gg
\Lambda_{\rm LO}$\footnote{The equality of
%similarity between 
$\alpha_s(M_Z^2)$ at LO and NLO approximations,
%and $\alpha^{\rm LO}_s(M_Z^2)$,
where $M_Z$ is the $Z$-boson mass, relates $\Lambda$ and $\Lambda_{\rm LO}$:
$\Lambda^{(4)} = 284$ MeV (as in \cite{ZEUS01}) corresponds to 
$\Lambda_{\rm LO} = 112$ MeV (see \cite{HT}).}
and, thus, $\alpha_s(Q^2)$ at LO is considerably smaller  then 
$\alpha_s(Q^2)$ at NLO  for HERA $Q^2$ values.

In other words, performing some resummation procedure (such as Grunberg's 
effective-charge method \cite{Grunberg}), one can see that the
%the NLO form 
results up to NLO approximation may
%can 
be represented as $\sim \alpha_s(Q^2_{\rm eff})$,
where $Q^2_{\rm eff} \gg Q^2$. 
Indeed, from 
different studies
\cite{DoShi,bfklp,Andersson},
it is well known that at small-$x$ values the effective
argument of the coupling constant is higher then $Q^2$.

Here, to improve the agreement at small $Q^2$ values without additional parameters,
we modify the QCD coupling constant.
We consider two modifications, 
which effectively increase the argument of the coupling constant 
at small $Q^2$ values (in agreement with \cite{DoShi,bfklp,Andersson}).

In one case, which is more phenomenological, we introduce freezing
of the coupling constant by changing its argument $Q^2 \to Q^2 + M^2_{\rho}$,
where $M_{\rho}$ is the $\rho $-meson mass (see \cite{Greco}). Thus, in the 
formulae of the
Section 2 we should do the following replacement:
\begin{equation}
 a_s(Q^2) \to a_{\rm fr}(Q^2) \equiv a_s(Q^2 + M^2_{\rho})
\label{Intro:2}
\end{equation}

The second possibility incorporates the Shirkov--Solovtsov idea 
\cite{ShiSo,Nesterenko,Cvetic}
about analyticity of the coupling constant that leads to the additional its
power dependence. Then, in the formulae of the previous section
%and \ref{Sec:3}
the coupling constant $a_s(Q^2)$ should be replaced as follows:
\begin{eqnarray}
 a^{\rm LO}_{\rm an}(Q^2) \, = \, a_s(Q^2) - \frac{1}{\beta_0}
 \frac{\Lambda^2_{\rm LO}}{Q^2 - \Lambda^2_{\rm LO}}
\label{an:LO} 
\end{eqnarray}
at the LO
%leading (LO) 
approximation and
\begin{eqnarray}
 a_{\rm an}(Q^2) \, = \, a_s(Q^2) - \frac{1}{2\beta_0}
 \frac{\Lambda^2}{Q^2 - \Lambda^2} 
+ \ldots \, ,
%- \frac{1}{\beta_0}
% \sum_{k=1}^\infty \left(\frac{\Lambda^2}{Q^2}\right)^k \, C_k[f]
\label{an:NLO}
\end{eqnarray}
at the NLO approximation,
where the symbol $\ldots$ stands for terms which have negligible
contributions
at $Q \geq 1$ GeV \cite{ShiSo}\footnote{Note that in \cite{Nesterenko,Cvetic} 
more accurate, but essentially more
cumbersome approximations of $a_{an}(Q^2)$ have been proposed.
We limit ourselves by above simple form (\ref{an:LO}), (\ref{an:NLO})
and plan to add the other modifications in our future investigations.}.

Figure~4 and Table 1 show  a strong improvement of the agreement with experimental data
for $F_2$ (almost 2 times!). Similar results can be seen also in Figs. 5 and 6 for
%shows 
the experimental data for $\lambda_{F_2}^{\rm eff}(x,Q^2)$
at $x\sim 10^{-3}$, which represents an average of the $x$-values of HERA experimental 
data. The top dashed line represents the aforementioned linear rise of
$\lambda(Q^2)$ with $\ln(Q^2)$.

%It 
So, Figures 5--7 demonstrate 
that the theoretical description of the small-$Q^2$ ZEUS
data for $\lambda^{\rm eff}_{F_2}(x,Q^2)$ by NLO QCD is significantly
improved by implementing the ``frozen'' and analytic coupling constants
$\alpha_{\rm fr}(Q^2)$ and $\alpha_{\rm an}(Q^2)$, 
respectively,
which in turn lead to
very close results (see also \cite{KoLiZo,Kotikov:2010bm}).

Indeed, the fits for $F_2(x,Q^2)$ in \cite{HT}
yielded
$Q^2_0 \approx 0.5$--$0.8$~GeV$^2$.
So, initially we had $\lambda^{\rm eff}_{F_2}(x,Q^2_0)=0$,
as suggested by Eq.~(\ref{1}). The replacements of Eqs.~(\ref{Intro:2}), (\ref{an:LO}) 
and (\ref{an:NLO}) modify the
value of $\lambda^{\rm eff}_{F_2}(x,Q^2_0)$. 
%So, for 
For the  
``frozen'' and analytic coupling constants 
$\alpha_{\rm fr}(Q^2)$ and $\alpha_{\rm an}(Q^2)$,
%coupling constant $a_{fr}(Q^2)$ 
the value of
$\lambda^{\rm eff}_{F_2}(x,Q^2_0)$ is nonzero 
%now 
and the slopes are
%is 
quite close to the experimental data at $Q^2 \approx 0.5$~GeV$^2$.
Nevertheless, for $Q^2 \leq 0.5$~GeV$^2$, there is still some disagreement with
the data, which needs additional investigation.
Note that at $Q^2 \geq 0.5$ GeV$^2$ our results are even better the results of 
phenomenological models \cite{CaKaMeTTV,Donnachie:2003cs}.

Figure~7 shows the $x$-dependence of the slope 
$\lambda^{\rm eff}_{F_2}(x,Q^2)$.
One observes good agreement between the experimental data and the generalized
DAS approach for a broad range of small-$x$ values.
%One can see an 
The absence of a variation with $x$ of 
$\lambda^{\rm eff}_{F_2}(x,Q^2)$ at small $Q^2$ values is related to the small
values of the variable $\rho$ there.

At large $Q^2$ values, the $x$-dependence of $\lambda^{\rm eff}_{F_2}(x,Q^2)$ is
rather strong.
However, it is well known that
the boundaries and mean values of the experimental $x$ ranges 
\cite{H1slo} increase proportionally with $Q^2$, which is related
to the kinematical restrictions in the HERA experiments:
$x \sim 10^{-4} \times  Q^2$
(see \cite{H197,ZEUS01,KoPa02}
and, for example,
Fig. 1 of \cite{Surrow}).
We show only the case with the ``frozen'' coupling constant because at large $Q^2$ values
all results are very similar.

From Fig.~7, one can see that HERA experimental data are close to
$\lambda^{\rm eff}_{F_2}(x,Q^2)$ at $x \sim 10^{-4} \div 10^{-5}$
for $Q^2=4$~GeV$^2$ and at $x \sim 10^{-2}$ for $Q^2=100$~GeV$^2$. Indeed,
%Incorporation of the
the correlations between $x$ and $Q^2$  in the form 
$x_{\rm eff}= a \times 10^{-4} \times Q^2$ with $a=0.1$ and $1$
%and $10$, 
lead
%in our %the analysis 
to a modification of the $Q^2$ evolution which starts
to resemble $\ln Q^2$, rather than $\ln \ln Q^2$ as is standard 
\cite{KoPa02}.\\
%It is in agreement with H1 data \cite{H197} for 
%$a$ values between $0.1$ and $1$ and $Q^2 > 2$ GeV$^2$, which approximately
%corresponds to the middle points of the measured $x$ range.

%{\bf Heavy quark part}\\
\section{Comparison with experimental data for SF $F_2^{cc}$ } \indent

We are now in a position to explore the phenomenological implications of our
results for SF $F_2^{cc}$.
As for our input parameters, we choose
%$Q_0^2=0.306$~GeV$^2$ \cite{Illarionov:2004nw},
$m_c=1.25$~GeV 
%and $m_b=4.7$~GeV 
in agreement with Particle Data Group
\cite{Amsler:2008zzb}.
While the LO result Eq.~(\ref{lo2}) is independent of the
unphysical mass scale $\mu$, the NLO formula~(\ref{eq:exp}) does depend on
it, due to an incomplete compensation of the $\mu$ dependence of $a_s(\mu)$ by
the terms proportional to $\ln(\mu^2/Q^2)$, the residual $\mu$ dependence
being formally beyond NLO.
In order to fix
%estimate
the theoretical uncertainty resulting from this, we put
$\mu^2=Q^2+4m_c^2$ (see (\ref{c_i})), which is the standart scale in heavy quark production.

\begin{figure}[t]
\begin{center}           
\includegraphics[width=\textwidth]{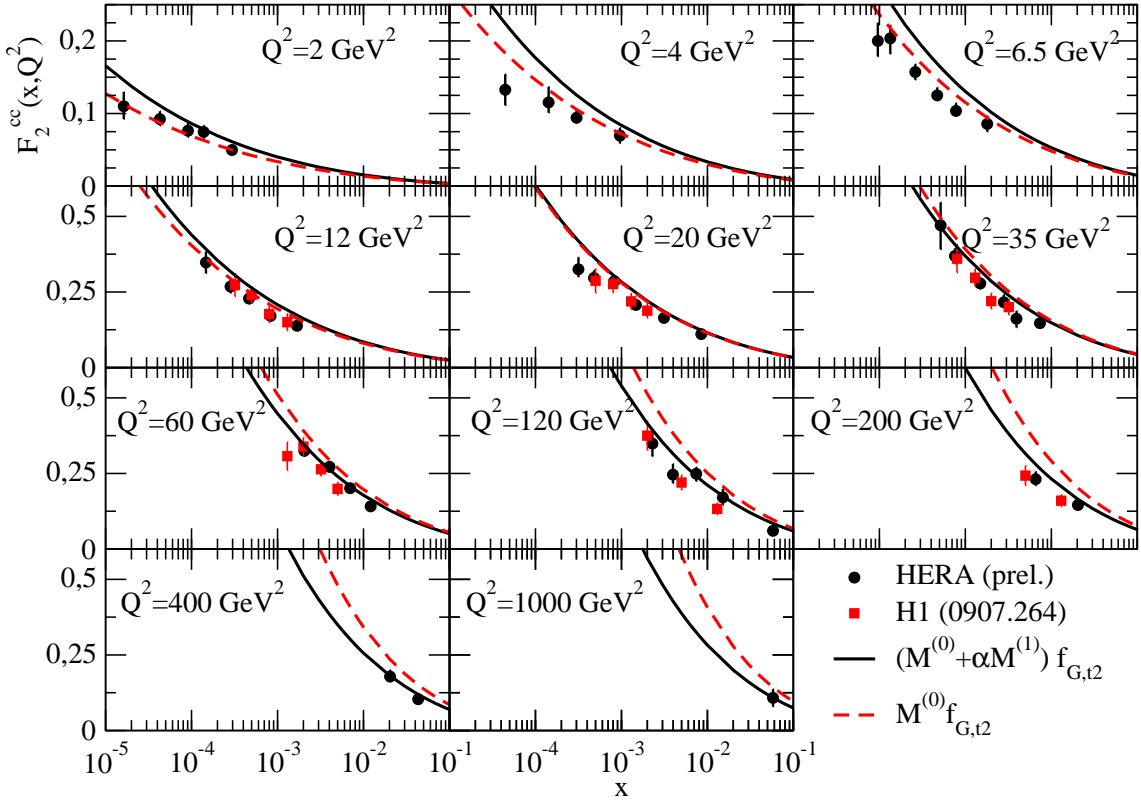}
\end{center}
\vskip 1.0cm
\caption{
$F_2^{cc}(x,Q^2)$ evaluated as functions of $x$ with the LO
matrix elements %from Eq. (\ref{lo2}) 
(dashed lines) and with the NLO ones %from Eq.~(\ref{nloA}) 
and
with the factorization/renormalization scale $\mu^2=Q^2+4m_c^2$ (solid lines).
The black points and red squares correspond to the the combine H1ZEUS
preliminary data \cite{Lipka:2009zza} and H1 data
 \cite{Collaboration:2009jy,:2009ut}, respectively.
}
\label{fig:r}
\end{figure}

The PDF parameters $\mu_0^2$, $A_q$ and $A_g$ shown in (\ref{1}), 
%(\ref{8.02}) and (\ref{8.02a}) 
have been fixed in the
fits of $F_2$ experimental data (see the previous section). Their values depend
%are performed to fix parameters of PDFs $xf_a(x,Q^2)$, which values depend
%usually for
on conditions chosen in the fits: the order of perturbation theory and the
number $f$ of active quarks.
%the structure function $F_2(x,Q^2)$

Below $b$-quark threshold,  the scheme with $f=4$ has been used
\cite{HT,Cvetic:2009kw} in the
fits of $F_2$ data.
% for $F_2$ structure function.
Note, that the $F_2$ structure
function contains $F_2^{cc}$ as a part. In the fits, the NLO gluon density
and the LO and NLO quark ones contribute to $F_2^c$, as the part of to $F_2$.
Then, now in PGF scattering the LO coefficient function (\ref{eq:lo})
corresponds in $m \to 0$ limit to the standart NLO Wilson coefficient
(together with the product of the LO anomalous dimension $\gamma_{qg}$ and
$\ln (m^2_c/Q^2)$. It is a general situation, i.e.
%. Moreover,
the coefficient funstion of
PGF scattering at some order of perturbation theory corresponds to the
standart DIS Wilson coefficient
%deep-inelastic scattering
with the one step  higher order.
%, which one step  higher. Indeed,
The reason is following:
the standart DIS
%deep-inelastic scattering
analysis starts with handbag
diagram of photon-quark scattering
and photon-gluon interaction begins
%can be introduces only
at one-loop level.

Thus, in our $F_2^{cc}$ analysis in the LO approximation of PGF process we
should take $f_a(x,Q^2)$
extracted from fits of $F_2$ data at $f=4$ and NLO approximation.
In practice, in \cite{Illarionov:2011km} we have applied our
$f=4$ NLO twist-two fit \cite{HT} of H1 data for $F_2$
with $Q^2$ cut: $Q^2> 1.5$ GeV$^2$, which produces $Q_0^2=0.523$~GeV$^2$,
$A_g=0.060$ and $A_q=0.844$.

Correspondingly, the NLO approximation of PGF process
%analysis
needs the gluon density exracted
from fits of $F_2$ data at NNLO approximation, which is not yet known
\footnote{The difficulty to extend the analysis
\cite{Q2evo,HT} to NNLO level is related with an
appearence of the pole $\sim 1/(n-1)^2$ in the three-loop corrections to the
anomalous dimension $\gamma_{gg}$ (see \cite{FaLi,Vogt:2004mw}). The pole
$\sim 1/(n-1)^2$ violates the Bessel-like solution (\ref{8.02}) of DGLAP
equation for
PDFs at low $x$ values with the flat initial condition (\ref{1}).}
in
generalized DAS
%double-asymptotic scalling 
regime. However, we see from the
modern global fits \cite{Dittmar:2005ed}, that the difference between NLO and
NNLO gluon
densities is not so large. So, we can apply the NLO form (\ref{8.02}) of
$f_a(x,Q^2)$ for our  NLO PGF analysis, too.

%Repeating all above discussions for the $Q^2$ range above the $b$-quark
%threshold, we conclude, that
%%Respectively,
%for $F_2^b$ analysis we should use  the NLO form (\ref{8.02}) of
%$xf_a(x,Q^2)$ with $f=5$.

The results for $F_2^{cc}$ 
%and $F_2^b$  
are prsented in Fig.8.
%respectively. From figures, we 
We can see a good agreement between our compact
formulas (\ref{eq:pm2n}), (\ref{eq:exp}), (\ref{lo2}) and (\ref{nloA})
an the modern experimental data \cite{Collaboration:2009jy}-\cite{Lipka:2009zza}
%\cite{Collaboration:2009jy,:2009ut,Abramowicz:2010zq,Chekanov:2009kj,Lipka:2009zza}
for $F_2^{cc}(x,Q^2)$ 
%and $F_2^b(x,Q^2)$ 
structure function.
%We would like to rememeber that the  agreement has been obtained
%without a free additional parameters.
To keep place on Fig.8, we show only the H1 \cite{Collaboration:2009jy,:2009ut}
data and the combine H1ZEUS preliminary \cite{Lipka:2009zza} one.

The good agreement between generalized 
double-asymptotic scaling 
DAS approach used here and $F_2$ and $F_2^{cc}$ 
%and $F_2^b$ 
data demonstrates an equal
importance of the both parton densities (gluon one and sea quark one) at low $x$.
It is due to the fact that $F_2$ relates mostly to the sea quark distribution, while the
$F_2^{cc}$  relates mostly to the gluon one. Dropping sea quarks in analyse ledas to the
different gluon densities extracted from $F_2$ of from $F_2^{cc}$
(see, for example, \cite{Jung:2007qh}).

\section{Conclusions} \indent

We have shown
%studied 
the $Q^2$-dependence of the structure functions $F_2$ and $F_2^{cc}$ and 
of the slope 
$\lambda^{\rm eff}_{F_2}=\partial \ln F_2/\partial \ln (1/x)$ at 
small-$x$ values in the 
framework of perturbative QCD. Our twist-two 
results are in very good agreement with 
%new 
precise HERA data at $Q^2 \geq 2$~GeV$^2$,
where perturbative theory can be applicable.
The application of the ``frozen'' and analytic coupling constants 
$\alpha_{\rm fr}(Q^2)$
and $\alpha_{\rm an}(Q^2)$ improves
%coupling constant $a_{fr}(Q^2)$
%(\ref{Intro:2}) 
%leads to good 
the agreement with the recent HERA data \cite{Surrow,H1slo,DIS02}
for the slope $\lambda^{\rm eff}_{F_2}(x,Q^2)$ for small $Q^2$ values,
$Q^2 \geq 0.5$~GeV$^2$.

%In this short paper, we 
We presented a compact formula for
%the ratio $R_i=F_L^i/F_2^i$
the heavy-flavour contributions to the proton structure functions $F_2$
valid through NLO at small values of Bjorken's $x$ variable.
Our results agree with modern experimental data
\cite{Collaboration:2009jy}-\cite{Lipka:2009zza}
%\cite{Collaboration:2009jy,:2009ut,Abramowicz:2010zq,Chekanov:2009kj,Lipka:2009zza}
%those extracted in Refs.~\cite{Aktas:2004az,Aktas:2005iw}
well within errors
without a free additional parameters.
In the $Q^2$ range probed by the HERA data, our NLO predictions agree very well
with the LO ones.
% and are rather stable under scale variations.
Since we worked in the fixed-flavour-number scheme, our results are bound to
break down for $Q^2\gg4m_i^2$, which manifests itself by appreciable QCD
correction factors and scale dependences.
As is well known, this problem is conveniently solved by adopting the
variable-flavour-number scheme, which not considered here.\\
%we leave for future work.\\

As a next step of investigations, we plan to perform combined fits of
the $H1\&ZEUS$ data \cite{Aaron:2009aa} of $F_2(x,Q^2)$, the $H1\&ZEUS$ data
\cite{Abramowicz:1900rp} of $F_2^{cc}(x,Q^2)$ and the
HERA data \cite{Collaboration:2009jy,Abramowicz:2010zq}
%\cite{H197,ZEUS01,Surrow,H1slo,DIS02}
of $F_2^{bb}(x,Q^2)$, using
% with 
the ``frozen'' and analytic 
coupling constants
%$\alpha_{fr}(Q^2)$ and $\alpha_{an}(Q^2)$
in both the LO and NLO approximations,
in order to improve the agreement with HERA data at small $Q^2$ values.
Several versions of the analytical coupling constant will be used.\\

\indent
%\vspace{1cm} \hspace{1cm} 
{\Large {} {\bf Acknowledgments} %\vspace{0.5cm}
}\\
\indent
This work was supported by 
RFBR grant 10-02-01259-a.
Author 
%A.V.K. 
thanks the Organizing Committee of 
XXI International Baldin Seminar on High Energy Physics Problems
for invitation and Paolo Bolzoni for discussions.
%\section{...}

\section{Appendix A}
\label{App:A}
\def\theequation{A\arabic{equation}}
\setcounter{equation}{0}

Here we give a short introduction of possible accounfing for BFKL corrections to
our analysis. As it was shown in Ref. \cite{Q2evo} for the first two orders 
of the perturbation theory, it is conveninet to start with Mellin moment
reprentation.

\subsection{Mellin moment form}

In the following we resume the steps we have followed to reach
the small $x$ approximate solution of DGLAP shown above (see also \cite{Q2evo})
\footnote{To work with BFKL formulas in the most symmetric way, in this
Appendix we will
%We 
use the normalization of the anomalous dimensions deviated by 
the factor ``$-1/2$'' from the DIS standard notation.}:
\begin{itemize}
\item Use the $n$-space exact solution for ``$\pm$''-components
\be
A_a^{\pm} \exp \left[- \int^{a_s(Q^2)}_{a_s(Q_{0}^2)} \, d\tilde{a}_s \,
\frac{\gamma_{\pm}(n,\tilde{a}_s)}{\beta(\tilde{a}_s)}\right]
\approx A_a^{\pm} e^{-d_{\pm}(n) s}
\ee
with
\be
d_{\pm}(n) = - \frac{\gamma^{(0)}_{\pm}(n)}{\beta_0}
\ee
\item Expand the perturbatively calculated parts
     (of anomalous dimensions  and coefficient functions) in the vicinity of the point $n=1$.
\item The singular part of the ``$+$''-component with the form (hereafter
 $n=1+\omega$)
 \begin{eqnarray}
A_a \omega^k e^{-\hat d s_{LO}/\omega} 
 \label{9.1}
 \end{eqnarray}
leads to modified Bessel functions in the $x$-space in the form 
\begin{eqnarray}
A_a {\biggl(\frac{\hat d s}{\ln x}\biggr)}^{(k+1)/2} 
I_{k+1}\biggl(2\sqrt{\hat d s \ln x}\biggr) 
 \label{9.2}
 \end{eqnarray}
\item The regular part $B(n) \exp{(-\overline d(n) s)}$ leads to the
additional coefficient (see Ref. \cite{Q2evo} and Appendix there)
$$B(1) e^{-\overline d(1) s} + O(\sqrt{\hat d s/ \ln x}) $$
 behind of the modified Bessel
function (\ref{9.2}) in the $x$-space. Because the accuracy is 
$O(\sqrt{\hat d_+ s/\ln x}) $,
it is necessary to use
only the basic term of Eq. (\ref{9.2}), i.e. all terms $\omega^k$ in front
of $\exp{(-\hat d_+/\omega)}$, with the exception of one with the
smaller $k$ value, can be neglected.
\item If the singular part at $n \to 1$ is absent (as in the case of the
``$-$''-component), i.e. $\hat d_- =0$ in
(\ref{9.1}), the result in the $x$-space is determined by 
$B(1) exp{(-\overline d(1) s)}$ with accuracy $O(x) $.
\end{itemize}

\subsection{BFKL corrections}
We would like to stress that the applicability of the above recipe
(to constract the small-$x$ solution which was shown in the previous 
subsection) is not limited by the order in perturbation theory but by the form
of the singular part of the anomalous dimensions.
At the first two orders of perturbation theory the singular part is
proportional to $\sim \omega^{-1}$
but this behaviour does not remain at higher orders.
The most singular terms
have been
calculated in \cite{BFKL}. For example, the singular part 
%of AD 
of the ``$+$''-component of the anomalous dimension matrix
%gluon-gluon AD 
has the following form
 \begin{eqnarray}
\gamma_{+}(\omega,a_s) = \gamma (\omega,a_s) +
O\biggl(a_s {\biggl(\frac{a_s }{\omega}\biggr)}^k \biggr) 
 \label{9.d}
 \end{eqnarray}
where
the terms $\sim O\Bigl(a_s (a_s /\omega)^k \Bigr) $
have been evaluated in Ref. \cite{FaLi}. 
%very recently.

The BFKL anomalous dimension $\gamma (\omega,\alpha)$ is obtained
by solving the implicit equation  
$$
1 ~=~ \frac{4C_A a_s }{\omega} \chi \Bigl( \gamma (\omega,a_s) \Bigr), 
$$
where the characteristic function $\chi ( \gamma )$ has the following
expression in terms 
of the Euler $\Psi$-function:
$$
\chi ( \gamma  ) ~=~ 2 \Psi(1) - \Psi (\gamma ) - \Psi (1- \gamma ),~~~
\Psi(\gamma)=\frac{d(\ln(\Gamma(\gamma))}{d\gamma}
$$

\subsubsection{Expansions}

The expansion of $\chi(n,\gamma)$ in powers of $\gamma$
%$a_s = 4 C_A \alpha $  
gives:
\begin{eqnarray}
\chi(\gamma) = \frac{1}{\gamma} + 2 \sum_{k=1}^{\infty} \zeta(2k+3) 
\gamma^{2k+2} \, \label{ex1}
%,\nonumber
\end{eqnarray}
that can be rewritten as the following exact relation for $\gamma$ itself
%is equal
\begin{eqnarray}
\gamma = \gamma_0 \left[1+  2 \sum_{k=1}^{\infty} \zeta(2k+3) 
\gamma^{2k+3}\right],~~ \gamma_0=\frac{4C_Aa_s}{\omega}\\
\, ,\nonumber
\end{eqnarray}
where $\gamma_0$ is the singular part of the LO gluon-gluon anomalous dimesion.
Solwing above eqution by interations, we have
%or 
%%approximately 
%equal to
\begin{eqnarray}
\gamma = \gamma_0 +   \sum_{k=3}^{\infty} C_k 
\gamma_0^{k+1}\nonumber 
\end{eqnarray}
%with the following nonzero $C_k$:
where the new coefficients are
\begin{eqnarray}
C_3=2\zeta(3),~ C_4=0,~ C_5=2\zeta(5),~ C_6=12\zeta^2(3),~ C_7=2\zeta(7), ...
\label{CK}
\end{eqnarray}

Incorporating the BFKL term to
%For 
the renormalization exponent leads to
%we have 
the following replacement
%additional contribution
\begin{eqnarray}
 \exp \{ -\frac{1}{\beta_0} \int^{a_s(Q^2)}_{a_s(Q_0^2)} 
\frac{da}{a^2}
\frac{\hat{\gamma}_{+}(a)}{\omega}\} 
%\nonumber \\
\to 
%\nonumber \\&& 
\exp \{ -\frac{1}{\beta_0} \int^{a_s(Q^2)}_{a_s(Q_0^2)} 
\frac{da}{a^2}\, \gamma(a)\}\nonumber
\label{Reno}
\end{eqnarray}
%and the following additional contribution\\

The contribution of the additional term $ \sum_{k=3}^{\infty} C_k 
\gamma_0^{k+1}$ in the r.h.s. of (\ref{CK}) has the following form
\begin{eqnarray}
 -\frac{1}{\beta_0} \int \frac{da}{a^2}\,
\sum_{k=3}^{\infty} C_k 
\gamma_0^{k+1}(a) = -\frac{1}{\beta_0} \frac{1}{a} \sum_{k=3}^{\infty} 
\frac{C_k}{k} 
\gamma_0^{k+1}(a)
\label{RenoN1}
\end{eqnarray}
%\vskip 0.5cm
and, thus, it has additional factor $k$ in the denominator. So, it gives a
hopeness
%So, there is the additional factor $k$ in denominator. We hope 
that
in the form (\ref{Reno}) the BFKL contributions
%corrections 
will be not so large as usual.

\subsubsection{Exact contribution}\

Now we considere the BFKL contribution to the r.h.s. of (\ref{Reno})
without any axpansions. Using integration by parts procedure, we obtain the
following expression

\begin{eqnarray}
\int \frac{da}{a^2}\, \gamma(a) = - \frac{1}{a}\, \gamma(a) +
\int \frac{da}{a}\, \frac{d\gamma(a)}{da}
= - \frac{1}{a}\, \gamma(a) + \int \frac{d\gamma}{a} \, ,
%\nonumber
\label{IBP}
\end{eqnarray}
where the inverse coupling constant is proportioanl to the characteristic 
function $\chi(\gamma)$ in (\ref{ex1})
\begin{eqnarray}
\frac{1}{a} = \frac{4C_A}{\omega} \, \chi(\gamma)\, ,
\nonumber
\end{eqnarray}
Thus, the last integral in (\ref{IBP}) can be evaluated exactly as
\begin{eqnarray}
\int \, d\gamma \, \chi(\gamma) = 2\Psi(1)\gamma + 
\ln \frac{\Gamma(1-\gamma)}{\Gamma(\gamma)}
\nonumber
\end{eqnarray}

So, the needed contribution in the r.h.s. of (\ref{Reno}) can be represented 
in the following form
%So, we have
\begin{eqnarray}
 \exp \left\{ -\frac{1}{\beta_0} \int^{a_s(Q^2)}_{a_s(Q_0^2)} 
\frac{da}{a^2}\, \gamma(a)\right\} = \frac{R(Q^2)}{R(Q_0^2)} \, ,
%\nonumber
\label{RenoN2}
\end{eqnarray}
where the $R$ value is (see also Ref. \cite{Kowalski:2012ur})
\begin{eqnarray}
R(Q^2) = {\left[\frac{\Gamma(\gamma)}{\Gamma(1-\gamma)}\right]}^{d_0}
\exp \left\{-d_0 \gamma \Bigl(\Psi(\gamma)+ \Psi(1-\gamma)\Bigr)\right\}
%\nonumber
\label{R}
\end{eqnarray}
with the new parameter $d_0$
\begin{eqnarray}
d_0 = \frac{4C_A}{\beta_0 \omega} = - \frac{\hat{d}_{+}}{\omega} = 
- \frac{\hat{d}_{gg}}{\omega}
\nonumber
\end{eqnarray}

When $a_s \to 0$ (i.e. in the considered case $\gamma \to 0$) we recover 
the singular part of the LO contributions (see the previous section)
 \begin{eqnarray}
R(Q^2) \to \gamma^{-d_0} \to a_s^{-d_0}
\nonumber
\end{eqnarray}

The transform of the Mellin moments in the form (\ref{RenoN2}) and (\ref{R})
to the Bjorken $x$-space, it is not a trivial problem. Author plans
to return to this problem in his future work.

\end{document}